\documentclass[11pt]{JHEP3}

\usepackage{amsfonts}
\usepackage{amsmath}
\usepackage{amsthm}
\usepackage{graphicx}

\def\braket#1{\mathinner{\langle{#1}\rangle}}
\newcommand{\sbraket}[1]{\lbrack #1\rbrack}

\newcommand{\boxit}[1]{%
  \[\fbox{%
      \addtolength{\linewidth}{-2\fboxsep}%
      \addtolength{\linewidth}{-2\fboxrule}%
      \begin{minipage}{\linewidth}%
      #1%
      \end{minipage}%
    } \nonumber \]%
}

\newcommand{\gz}{\zeta}
\newcommand{\gt}{\theta}

\newcommand{\cF}{\mathcal F}

\newcommand{\cK}{\mathcal K}

\newcommand{\cO}{\mathcal O}

\newcommand{\bC}{\mathbb C}

\renewcommand{\Im}{\mbox{Im~}}

\renewcommand{\Re}{\mbox{Re~}}

\newcommand{\be}{\begin{equation}}
\newcommand{\bea}{\begin{eqnarray}}

\newcommand{\ee}{\end{equation}}
\newcommand{\eea}{\end{eqnarray}}
\newcommand{\half}{\frac{1}{2}}

\newcommand{\ov}[1]{\frac{1}{#1}}

\newcommand{\dbar}{\bar\partial}

\newcommand{\ii}{\textrm{i}}
\newcommand{\al}{{\alpha'}}

\newcommand{\sa}{\sqrt{2\alpha'}}

\theoremstyle{definition}

\title{On-shell Recursion in String Theory}
\author{Rutger H. Boels \\ Niels Bohr International Academy and DISCOVERY center, Niels Bohr Institute,  Blegdamsvej 17, DK-2100 Copenhagen, Denmark}
\author{Daniele Marmiroli and Niels A. Obers \\ Niels Bohr Institute, Blegdamsvej 17, DK-2100 Copenhagen, Denmark }


\abstract{We prove that all open string theory disc amplitudes in a flat background obey Britto-Cachazo-Feng-Witten (BCFW) on-shell recursion relations, up to a possible reality condition on a kinematic invariant. Arguments that the same holds for tree level closed string amplitudes are given as well. Non-adjacent BCFW-shifts are related to adjacent shifts through monodromy relations for which we provide a novel CFT based derivation. All possible recursion relations are related by old-fashioned string duality. The field theory limit of the analysis for amplitudes involving gluons is explicitly shown to be smooth for both the bosonic string as well as the superstring. In addition to a proof a less rigorous but more powerful argument based on the underlying CFT is presented which suggests that the technique may extend to a much more general setting in string theory. This is illustrated by a discussion of the open string in a constant B-field background and the closed string on the level of the sphere.}

\begin{document}

\section{Introduction}
Recent years have seen many new techniques and insights for the calculation of scattering amplitudes within field theory inspired by Witten's twistor string proposal \cite{Witten:2003nn}. Since field theory arises as the low energy limit of ordinary string theory in a flat background, a natural question is to what extent these advances for fields can be carried over to strings in this setting. This is especially important as much useful information about the string theory can be obtained directly from the amplitudes and vice-versa. In this context it is noteworthy that much of the recent field theory progress is based on the same `analytic S-matrix' type approach which originally led to the birth of string theory \cite{Veneziano:1968yb}. Besides aiming for direct benefits in field theory applications there is an intrinsic interest in understanding string theory better. Although in this article the focus will be mostly on flat backgrounds, a longer term goal of our program is to understand string theory in curved backgrounds as this relates directly to strong coupling field theory through the AdS/CFT correspondence \cite{Maldacena:1997re}.

In this article the Britto-Cachazo-Feng-Witten (BCFW) on-shell recursion relations \cite{Britto:2004ap,Britto:2005fq} are studied in the context of string theory. In field theory these useful relations relate tree level amplitudes to a sum over amplitudes with a smaller number of particles, evaluated at complex values of the momenta. The elementary and elegant derivation of the relations \cite{Britto:2005fq} involves a complex momentum shift on two particles. Crucially, an absence of certain residues at infinite momentum shifts needs to be shown to make the relations work. Residues at infinite complex momentum are at the least by hand waving related to the UV behavior of the theory under study. This reasoning can be made more precise in field theory \cite{Forde:2007mi}. Absence of residues at infinity has been proven in (super)Yang-Mills and Einstein (super)gravity in any dimension from four onwards, see \cite{ArkaniHamed:2008gz}, \cite{Cheung:2008dn} and references therein. Of course, string theory has excellent UV behavior and a natural question is whether string theory amplitudes also obey similar on-shell recursion relations.

The string version of on-shell recursion relations was first investigated by K. Larsen, M. Vonk and two of the present authors in \cite{Boels:2008fc}, where these relations were shown to hold for all four-particle amplitudes in both the open and the closed string in a flat background and for a certain five-particle amplitude in four dimensions. Already there it was striking how natural on-shell recursion appears in string theory.  However, the analysis of \cite{Boels:2008fc} was based on fully integrated forms of the amplitude. As these are unknown for larger ($> \sim 5$) numbers of external particles\footnote{To be precise, an explicit expression for the integrated five-point vector amplitude in the superstring was first obtained in \cite{Medina:2002nk}, later simplified in \cite{Barreiro:2005hv} and more recently in \cite{Mafra:2009bz}. In \cite{Oprisa:2005wu} the general six-point vector amplitude was written explicitly in terms of $6$ basis integrals which are amenable to an explicit $\al$ expansion (also developed there). In \cite{Stieberger:2007jv} a similar form of the MHV amplitude in four dimensions for seven vector particles was obtained.}, the approach of \cite{Boels:2008fc} is unsuitable for a general analysis. The aim of this paper is to change this situation.

An outline and summary of the main results is as follows. We start in Section \ref{sec:proofopenstringrecadj} by examining adjacent shifts in open string amplitudes based on the integral expressions, up to the determination of a certain kinematic constraint. This constraint is determined explicitly for amplitudes with shifted gluon legs in both the bosonic string as well as the superstring. Non-adjacent shifts are treated in Section \ref{sec:nonadjshift}, based on the monodromy relations for string theory amplitudes first presented in \cite{Plahte:1970wy} and recently discussed in \cite{BjerrumBohr:2009rd} and \cite{Stieberger:2009hq}. It is shown how these monodromy relations follow directly from the underlying CFT which may be of independent interest. The outlines of a completely CFT-based derivation of the adjacent shift behavior of any amplitude is presented in Section \ref{sec:CFTderivation}. This powerful technique will be applied in two sample cases in Section \ref{sec:curvback}. One is a generalization of our results to open string amplitudes in constant B-field backgrounds. The other is a CFT analysis of the large shift behavior for the closed string. An independent argument based on the Kawai-Lewellen-Tye (KLT) relations \cite{Kawai:1985xq} is also included there. A discussion, conclusions and some speculations  round off the main presentation in Section \ref{sec:conc}. In Appendix \ref{fourgluons} shifts of the four point gluon amplitude in the bosonic string and the superstring are presented as a worked-out example. Appendix \ref{app:antisymsuperstring} contains a proof of a technical point. Finally, in Appendix \ref{app:CSWshifts} naive application of a three particle shift originally used to derive the CSW rules in field theory is shown to lead to inconsistent results for four particle superstring amplitudes.

\textit{Note added in proof}: While this paper was being readied for publication \cite{Cheung:2010vn} appeared which has a sizeable overlap with the techniques and results presented here, especially in section \ref{sec:CFTderivation}.

\section{Recursion relations for the disc: adjacent shifts}
\label{sec:proofopenstringrecadj}
\subsection{Lightning review of on-shell recursion}
The key observation for deriving the on-shell recursion relations is
that any tree level scattering amplitude can easily be turned into a
rational function of a single complex variable by deforming the momenta
\cite{Britto:2005fq}, requiring that these deformed momenta remain on-shell and obey
momentum conservation. The simplest example of this is to take two
particles $i$ and $j$ and shift their momenta by a vector $q_{\mu}$
\begin{align}
p^{\mu}_i &\rightarrow \hat{p}^{\mu}_i = p_i^{\mu} + z q^{\mu} \ , \nonumber \\
p^{\mu}_j &\rightarrow \hat{p}^{\mu}_j = p_j^{\mu} - z q^{\mu} \ ,
\label{eq:genBCFWshift}
\end{align}
which preserves momentum conservation. For an example of a more complicated shift see  appendix
\ref{app:CSWshifts}. For two particle shifts linear in $z$ as in eq.~\eqref{eq:genBCFWshift}, the on-shell constraint is satisfied iff the vector $q$ obeys
\begin{equation} \label{eq:constraintnmu}
p^{\mu}_i q_{\mu} = p^{\mu}_j q_{\mu} = q^{\mu} q_{\mu} = 0 \ .
\end{equation}
These equations do not have a solution for real $q_\mu$, but do for
complex momenta, as can easily be verified by going to the common lightcone frame.

After the shift any $n$-point amplitude $A_n$ becomes a function of a complex variable $A_n(z)$,
where the amplitude of interest is of course $A_n(z=0)$. This can be obtained by an
elementary contour integration around a contour which only
encompasses the pole at $z=0$,
\begin{equation}
\label{eq:BCFWstarting} A_n (0) = \oint_{z=0} \frac{A_n(z)}{z} d z \
.
\end{equation}
If the contour is now pulled to the other side of the Riemann sphere
one encounters various poles at finite values of $z$ and a possible
residue at infinity,
\begin{equation}\label{eq:pullingcontourtoinf}
A_n (0) = \oint_{z=0} \frac{A_n(z)}{z} dz = - \left\{\sum
\mathrm{Res}_{z=\textrm{finite}} + \mathrm{Res}_{z= \infty}
\right\} \ .
\end{equation}
The poles at finite values of $z$ correspond to the exchange of
physical particles. By tree level unitarity, the residues at these
poles must be the product of two tree level amplitudes with each one leg
containing the particle being exchanged, summed over all particles
at this particular mass level. The residue at infinity does not have a similar physical interpretation.  \emph{If} therefore this residue vanishes then all terms on the right hand side of \eqref{eq:pullingcontourtoinf} are known and consist of lower point
amplitudes. Therefore in this case a recursion relation is obtained
between amplitudes
\begin{equation}
A_n(1,2,3 \ldots, n) = \sum_{r,h(r)} \sum_{k=2}^{n-2}
\frac{A_{k+1}(1,2,\ldots, \hat{i}, \ldots, k, \hat{P}_r)
A_{n-k+1}(\hat{P}_r, k+1, \ldots, \hat{j}, \ldots, n)}{\left(p_1 +
p_2 + \ldots + p_k \right)^2 +  m_r^2} \ , \label{eq:recursiongen}
\end{equation}
where the first sum is over all different mass levels $r$ and over all
polarization states at that level, denoted $h(r)$. The momentum
$\hat{P}_r$ for the `extra' particle and its anti-particle in the
amplitude is such that the particle is on-shell. This condition determines the numerical value of $z$. The second sum over $k$ is over all the different ways in which the amplitude can be factorized with the shifted legs on the different amplitudes in the residues. Note that for every different term in the $k$ sum the numerical value of $z$ 
entering the residue is different.

The challenge in deriving this relation is proof of absence of the residue at infinity. This proof will be provided for all open string theory amplitudes in a flat background in this paper, subject to a kinematic constraint. As the derivation of the BCFW recursion relation involves a limit, the field theory limit of the resulting equations has to be treated with care \cite{Boels:2008fc} to avoid `order of limits' problems. Note that any symmetry of the three-point amplitude will imply through the recursion relations a corresponding symmetry of $n$-point amplitudes.

\subsubsection*{Dimensionality of space-time}
Our analysis of on-shell recursion in string theory will hold in principle for any dimensionality of the target space-time. However, in the recursion relations one has to sum over \emph{all} particles appearing in the theory. Hence in dimensions above the usual critical dimension of the string theory of interest negative norm states have to be included for instance. In dimensions below the critical dimension, one can use a dimensional reduction argument to reduce the amplitude from the critical dimension to fields in the dimension one is interested in. Although it would certainly be interesting to study non-critical string amplitudes this is beyond the scope of the present paper.

\subsubsection*{Relation to factorization formulae}
The recursion relation in eq.~\eqref{eq:recursiongen} has a passing resemblance to factorization formulas as studied in the beginning days of string theory. See for example equation $90$ in \cite{DiVecchia:2007vd}, or chapter 7 of \cite{Green:1987sp}. One can write any string amplitude in terms of so-called Feynman-like diagrams. Diagrammatically, this corresponds to molding the string world-sheet into the rough shape of a particular Feynman graph with a certain manifest singularity structure. In formulas this reads,
\begin{equation}
A = \langle k_1 | V_2 D V_3 \ldots V_{n-1}| k_n \rangle \ , 
\end{equation}
where $D$ is the string theory propagator $\frac{1}{L_0 -1}$. Now one can insert a complete set of states next to one propagator, say the first one,
\begin{equation}\label{eq:facform}
A = \langle k_1 | V_2 \left(\sum_{\lambda} |\lambda, P  \rangle \langle P, \lambda | \right) \left(\frac{1}{L_0 -1} \right) \left(\sum_{\lambda'} |\lambda', P  \rangle \langle P, \lambda' | \right) V_3 \ldots V_{n-1}| k_n \rangle \ , \end{equation}
where
\begin{equation}
P= k_1 + k_2 \ . 
\end{equation}
This expression has manifest poles as a function of $P$: these are such that the states are annihilated by $L_0-1$. Therefore, precisely at these poles the amplitude factorizes. This is one way to see that string theory amplitudes have the poles required by tree level unitarity. Away from the poles however one has to sum over all states in the Hilbert space of the harmonic oscillators of the string which include unphysical modes. Demonstrating that these modes decouple in physical amplitudes (i.e. in the residue at the pole) was the main objective of the beginning days of string theory. In general however, the momentum in the channel for which the poles are displayed is not on-shell. Hence eq.~\eqref{eq:facform} certainly does not express an amplitude in terms of lower point \emph{amplitudes}, except in the limit where the sum of the momenta on the left hand side squared approaches its pole value. In contrast,  \eqref{eq:recursiongen} expresses the amplitude for generic momenta as a sum over a subset of the poles of the amplitude, with the residue modified by the shifted momenta.

One can point to more differences between factorization formulas and on-shell recursion. For instance, the formula in eq.~\eqref{eq:recursiongen} is a sum over all the channels for which the shifted momenta appear on the left and right hand side, whereas eq.~\eqref{eq:facform} above displays poles in one channel only. Moreover, eq.~\eqref{eq:recursiongen} involves shifted momenta. That implies for one that the numerical values of the momenta on the amplitudes for both the shifted legs as well as the intermediate channel are different from the above factorization formulae and are different for every different channel in the sum.

\subsection{Veneziano revisited}
To get an idea for how to proceed, it is instructive to revisit the (ordered) Veneziano amplitude \cite{Veneziano:1968yb} for the scattering of four tachyons in open bosonic string theory,
\begin{equation}\label{eq:venezianoamp}
A(s,t) = \frac{\Gamma(\al s-1) \Gamma(\al t -1)}{\Gamma(\al(s+t)
-2)} \ ,
\end{equation}
where $s$,$t$ and $u$ are the usual Mandelstam variables,
\begin{equation}
s = (p_1 + p_2)^2 \quad t = (p_1 + p_4)^2 \quad u = (p_1 + p_3)^2 \
.
\end{equation}
We remind the reader that the full amplitude is the sum over non-cyclic orderings of the above expression, possibly dressed with Chan-Paton factors. In the following all open string amplitudes will be considered to be color ordered. There are three different BCFW-type shifts possible of \eqref{eq:venezianoamp}: two of adjacent particles and one of non-adjacent ones. These have been discussed in detail in \cite{Boels:2008fc} using the properties of the Gamma function and are related to Regge behavior of the four particle amplitude.

Inspired by the close analogy to Regge behavior and the analysis of this for the five-point amplitude in \cite{Bardakci:1969mg} in this paper the BCFW shift will be studied directly from the well-known integral representation,
\begin{equation}\label{eq:venezianoampintegral}
A(s,t) = \int_{0}^1 d\,y y^{\al s - 2} \left(1-y\right)^{\al t - 2}
\ .
\end{equation}
For concreteness, consider the shift for particles $1$ and $4$ for which
\begin{equation}\label{eq:shiftmandvar}
\al \hat{s} = \al s + z' \quad \al \hat{t} = \al t \quad \al \hat{u}
= \al u - z' \ .
\end{equation}
holds. Here the change of variables $z' = 2 \al z (p^{\mu}_2 q_{\mu})$ has
been employed the contour integral. To study the residue at infinity
of the resulting expression, the corresponding integral in
\eqref{eq:venezianoampintegral} must be evaluated in the limit of
large $z'$ in any direction of the complex plane. In the above expression the
following change of variables is useful
\begin{equation}
y = \exp\left(- \frac{\beta w}{\al s - 2 + z'} \right) \ ,
\end{equation}
to transform the integral to
\begin{equation}
A(z') = \int_{0}^{\infty} \frac{\beta dw}{\al s -2 + z'}
e^{-\frac{\beta w}{\al s-2+z'}}
 \left( 1-e^{- \frac{\beta w}{\al s -
2 + z'}}\right)^{\al t - 2} e^{-\beta w} \ .
\end{equation}
The boundary values of this integral are correct as long as
\begin{equation}
\Re \left(\frac{\beta w}{\al s - 2 + z'} \right) > 0  \ .
\end{equation}
If $z'$ is taken to $-\infty$ along the real axis with $\Re(\beta)>0$, it is easy to see that the amplitude can be expanded
as a non-holomorphic function times a Laurent series,
\begin{eqnarray}
A(z') = & &  - \int_{0}^{\infty}  \beta dw  \frac{\beta}{z'} \left(-\beta w /z' \right)^{\al t - 2} e^{- \beta w}  \nonumber \\
&& \times \left(1 + \frac{1}{z'} \left(2-\al s +  \al t (\al s -2 +
\frac{1}{2} \beta w) \right) +
\mathcal{O}\left(\frac{1}{z'^2}\right) \right) \ .
\end{eqnarray}
If
\begin{equation}
\Re \left(\al (p_1+p_4)^2 \right) > 1 \qquad {\rm and} \qquad
  \Re \left(\beta \right) > 0 \ ,
\end{equation}
the resulting $w$ integral can be performed to yield
\begin{equation}\label{eq:openstringregge4pt}
A(z') \rightarrow \left(-\frac{1}{z'}\right)^{\al t -1}
\Gamma\left(\al t - 1 \right) \left(1 + \frac{1}{z'}(\al t -1)(\al s
- 2 + \frac{1}{2} \al t ) + \mathcal{O}\left(\frac{1}{z'^2}\right)
\right) \ .
\end{equation}
From the analysis it is clear that for every ray in the complex $z'$
plane apart from the positive real axis there is a $\beta$ for which
$\Re(\beta)>0$ such that the amplitude will behave like
\eqref{eq:openstringregge4pt}. The integral of the resulting function
around a large contour with a point excised on the real axis then
vanishes as long as the kinematic constraint
\begin{equation}\label{eq:realityconstraintveneziano}
\Re \left(\al (p_1+p_4)^2 \right) > 1  \ ,
\end{equation}
is satisfied.
This reproduces the result of \cite{Boels:2008fc} through a direct
integral derivation. To complete the proof of BCFW recursion the
excised region must be examined: an infinitesimal contour segment
which intersects the positive real axis. Since the function under study is
analytic on this line segment, the (absolute value of the) resulting
integral vanishes as the length of the segment is taken to zero.

\subsubsection*{BCFW shifts versus essential singularities}
The above argument about large $z$ behavior might seem confusing in
the light of the fact that the Beta function in the integrated form
of the Veneziano amplitude \eqref{eq:venezianoamp} is known to have
an essential singularity if one of the arguments is taken to
infinity. This seems in sharp contrast with the behavior for the
residue at infinity derived above. The resolution of this point is that the integral contours have to be defined with care. To excise an infinitesimal arc for instance the function under study has to be analytic. This is only true away from the poles in the Gamma function. Similarly, it is assumed implicitly when writing the infinite sum that the contour integrals are well-defined. Both these implicit assumptions are violated if the contours are chosen to be limiting towards the poles, as this is
where the essential singularity is located. If the contours are chosen to avoid the poles then no problem arises.

To illustrate this point in the above example, for the Veneziano amplitude a good choice
of contours are circles of radius $R_k$,
\begin{equation}
R_k=k+\frac{1}{2} \quad \quad k \in \mathbb{N} \ .
\end{equation}
A bad choice of contours would be circles with radius $R'_k$
\begin{equation}
R'_k=k+\frac{1}{k} \quad \quad k \in \mathbb{N} \ .
\end{equation}
as these limit toward the poles of the amplitude.

\subsection{Adjacent shifts for all multiplicities}
\subsubsection{Tachyon amplitudes in the bosonic string}
The analysis for adjacent shifts of the Veneziano amplitude above can be generalized to the open string tachyon scattering amplitude for all multiplicities which are given by the well-known  Koba-Nielsen formula
\cite{Koba:1969rw},
\begin{equation}
A_n = \int_{0 \leq y_{n-1} \leq \ldots \leq y_{3}\leq 1}
\prod_{2 < i<j < n}\left(y_i - y_j\right)^{2 \al p_i p_j} \ .
\end{equation}
In deriving this expression from the path integral the positions of $3$ vertex operators have been
fixed: particles $1$, $2$ and $n$ at $\infty$, $1$ and $0$
respectively. Despite appearances, this expression can be shown to
be cyclically symmetric in the external legs. We can therefore shift
any two adjacent particles to cover all adjacent shifts and for the
above expression it is convenient to choose $n$ and $1$. Through a
coordinate transformation (see e.g. the useful review \cite{DiVecchia:2007vd})
\begin{equation}
\label{udef} u_i = \frac{y_{i+1}}{y_i} \quad, \quad 2 \leq i \leq n-2 \ ,
\end{equation}
this expression can be transformed to
\begin{equation}\label{eq:handyforshifts}
A_n = \left(\prod_{i=2}^{n-2} \int_0^1 du_i u_i^{\al s_i -2} \right)
\left(\prod_{k=2}^{n-2} \prod_{j=k+1}^{n-1} \left(1-\prod_{l=k}^{j-1}
u_l \right)^{2 \al p_k p_j}\right) \ ,
\end{equation}
with $s_i = (\sum_{k=1}^i p_i)^2$. From this expression it is easy
to see that when particles $1$ and $n$ are shifted one can apply the
integral argument given above for the Veneziano amplitude several
times to obtain the limiting behavior for $z \rightarrow \infty$.

Concretely, the chosen shift shifts
\begin{equation}\label{eq:shiftnparticles}
s_i \rightarrow \hat{s}_i =  s_i + 2 z q_{\mu} \left(\sum_{k=2}^i
p^{\mu}_i \right) \equiv s_i + \frac{\gamma_i}{\al} z \ .
\end{equation}
In line with the analysis above, change coordinates to
\begin{equation}
\label{uwchange} u_i = \exp \left(- \frac{\beta_i w_i}{\al s_i -2 + \gamma_i z } \right) \equiv e^{- \tilde{w}_i} \ ,
\end{equation}
which turns \eqref{eq:handyforshifts} into
\begin{equation}\label{eq:tachbosshifteq}
A_n(z) = \left(\prod_{i=2}^{n-2} \int_0^\infty dw_i \left(
\frac{-\beta_i e^{- \tilde{w}_i}}{\al s_i +  \gamma_i z - 2} \right)
e^{- \beta_i w_i}  \right)  \left(\prod_{k=2}^{n-2}
\prod_{j=k+1}^{n-1} (1-e^{-\sum_{l=k}^{j-1} \tilde{w}_l})^{2 \al p_k
p_j}\right) \, ,
\end{equation}
accompanied by the reality conditions
\begin{equation}
\Re \left(\frac{\beta_i w_i}{\al s_i -2 + \gamma_i z} \right) > 0
\quad  , \quad \ \Re \left(\beta_i \right) > 0 \ .
\end{equation}

From eq.~\eqref{eq:tachbosshifteq} the large $z$ behavior of the bosonic string tachyon amplitude follows as
\boxit{\begin{equation}
A_n(z) \sim \left(\frac{1}{z}\right)^{\al (p_1 + p_n)^2 -1}
\left(G_0 + \frac{G_1}{z} +
\mathcal{O}\left(\frac{1}{z}\right)^2\right) \ ,
\end{equation}}
which is the result from a Laurent expansion around $z=\infty$. In this expression $G_i$ denote certain $(n-3)$-fold exponential integrals that we have not been able to integrate exactly, but which do not depend on $z$.
This is sufficient for our purposes as the above form completely
isolates the large $z$ behavior of the complete amplitude. Using a
similar analysis as above, we conclude that for adjacent shifts the
Koba-Nielsen amplitude obeys BCFW recursion if
\begin{equation}\label{eq:tachyonreccondition}
\Re \left(\al (p_{i}+p_{i+1})^2 \right) > 1  \ ,
\end{equation}
with $i$ and $i+1$ the labels of the shifted particles. More
precisely, $n-3$ points on the contour integral must be excised. It
can then be argued that their contribution vanishes because of
analyticity of the integrand on the contour. In principle $G_0$
could integrate to zero, so the above analysis establishes
a bound only. Vanishing coefficients might be a signal of an underlying
symmetry.

\subsubsection*{Gluon amplitudes in the bosonic and super cases}
\label{gluonbosonic}
The main difference of the tachyon amplitudes and amplitudes involving other modes of the string are the complications caused by polarization vectors as these must be transverse to the shifted momenta. As a concrete and important example of this, adjacent shifts of the general $n$-point gluon amplitudes in the bosonic and superstring will be considered in this subsection. To solve the complication and obtain concrete expressions it is instructive as was done in \cite{ArkaniHamed:2008yf} to consider the lightcone frame of the two shifted momenta,
\begin{equation}
p_1 = \frac{1}{\sqrt{2}} (1,1,0,0;\dots0) \ , \qquad p_n = \frac{1}{\sqrt{2}} (1,-1,0,0;\dots0) \ ,
\end{equation}
where we have set the energy scale by one of the momenta to avoid cluttering formulas later. In this frame the shift vector obeying \eqref{eq:constraintnmu} can be chosen to be
\begin{equation}
q= \frac{1}{\sqrt{2}} (0,0,1,i;\dots0) \ .
\end{equation}
With this shift choice it is convenient to choose the polarization vectors for unshifted momenta as
\begin{equation}
\gz_1^-=\gz_n^+=q \ , \qquad \gz_1^+=\gz_n^-=q^* \ , \qquad \gz^T=(0,0,0,0;\dots,1,\dots,0) \ .
\end{equation}
These vectors are given in a lightcone gauge in which the lightcone gauge vector of one leg is the momentum of the other leg. Under a momentum shift
\begin{equation}
p_1 \rightarrow p_1+qz \ , \qquad p_n \rightarrow p_n-qz  \ ,
\end{equation}
the set of transformations that leaves the transversality constraint $\gz_i\cdot p_i =0$ invariant reads
\begin{equation}
\label{shift_rules}
\begin{array}{ccccccc}
\gz_1^- & = & \gz_n^+ & = & q   & \rightarrow & q\\
        &   & \gz_1^+ & = & q^* & \rightarrow & q^*+zp_n \\
        &   & \gz_n^- & = & q^* & \rightarrow & q^*-zp_1 \\
        &   &         &   & \gz^T & \rightarrow &\gz^T
\end{array}  \ .
\end{equation}
In fact, this can all be phrased covariantly by employing the higher dimensional spinor helicity method developed in \cite{Boels:2009bv}. In the above the polarization vectors are in the gauge in which the momentum of the other leg is the lightcone gauge vector. It is easily seen that the space of shift vectors is therefore spanned by the gluon polarization vectors described in \cite{Boels:2009bv}, with the vector `$q$' of that reference identified with the momentum of the other leg. The vector `$q$' above is identified with one of the polarization vectors in \cite{Boels:2009bv}. The BCFW shift in this setup amounts to shifting the pure spinors.

It will be found below that the structure of the argument is remarkably similar in form to the argument in \cite{ArkaniHamed:2008yf}. There it was shown that for adjacent shifts of amplitudes in Yang-Mills theory
\boxit{\begin{equation}\label{eq:largezfieldt}
A_n(z) \sim \hat{\gz}_1^{\mu}  \left(z \, \eta_{\mu\nu} h_1\left(\ov{z}\right)  + B_{\mu\nu} h_2\left(\ov{z}\right)  + \mathcal{O}\left(\frac{1}{z}\right)  \right) \hat{\gz_2}^{\nu}  \ ,
\end{equation}}
with some polynomial functions $h_i(\ov{z})$ for which $h_i(0)$ is a non-trivial constant. The matrix $B$ is anti-symmetric. To discuss behavior under shifts it is useful to note that the on-shell Ward identity for a shifted gluon leg reads
\be
\label{eq:wardid} \hat{p}_1^\mu
V_{\mu\nu}(z)=0 \quad \rightarrow \quad q^\mu V_{\mu\nu}(z)=-\ov{z}p_1^\mu
V_{\mu\nu}(z) \ .
\ee
This can be used to lower the power of $z$ by $1$ for those polarizations in \eqref{shift_rules} which are proportional to $q$. Working through the dependence of the external polarization vector on $z$ as was done in \cite{ArkaniHamed:2008yf} now yields Table \ref{tab:kinconfield}. The difference between the transversal polarizations T2 and T is whether or not $\gz_1^T \cdot \gz_n^T=0$  respectively. Below the analogue of this table for bosonic string theory and superstring theory will be presented.

\begin{table}
\begin{center}
\begin{tabular}{c|c c c}
$\gz_1 \;\backslash \;\gz_n  $ & $-$              & $+$              & T \\
\hline
$-$                   & $ +1$ & $ +1$ & $ +1$ \\
$+$                   & $ -3$ & $ +1$ & $ -1$ \\
T                     & $ -1$ & $ +1$ & $ -1$ \\
T2                    & $ -1$ & $ +1$ & $0$ \\
\end{tabular}
\caption{The leading power in $z^{-\kappa}$ for large $z$ limit of the adjacent shift of an all gluon amplitude in \textbf{field theory} for all possible polarizations \cite{ArkaniHamed:2008yf}.
  \label{tab:kinconfield}}
\end{center}
\end{table}

\subsubsection{Gluon amplitudes in the bosonic string}
The $n$-point gluon amplitude in the bosonic string is readily computed by exponentiating the polarization dependence into the vertex operator,
\begin{equation}
V(y,\gz,p)=\exp(i p \cdot X + \frac{1}{\sa}\gz \cdot \dot X)  \ ,
\end{equation}
with the prescription that only the multi-linear part in each of the polarization vectors arising from the expansion of the exponential has a physical meaning \cite{Green:1987sp}. For this vertex operator to have the right conformal dimensions,
\begin{equation}\label{eq:confconstraints}
p_i^2 = \gz_i\cdot p_i =0  \ ,
\end{equation}
must hold, while polarization vectors are identified under arbitrary shifts by the momentum,
\begin{equation}
\gz_i \rightarrow \gz_i + f p_i  \ ,
\end{equation}
for some function $f$.

The $n$-point gluon amplitude in the open bosonic string reads
\begin{equation}
\begin{split}
\label{eq:boson_vector} A_n = (y_A^0 - y_B^0)(y_B^0 - y_C^0)(y_A^0 - y_C^0)
\int_{\Omega} \prod_{i=1}^{n}dy_i \,\, \delta(y_A-y_A^0)
\delta(y_B-y_B^0)\delta(y_C-y_C^0)\times \\ \times \prod_{i<j}
(y_i-y_j)^{2 \al p_i \cdot p_j} \cF(y_i,\zeta_i,p_i) \ ,
\end{split}
\end{equation}
where $\Omega=\{y_1 \geq y_2 \geq \dots \geq y_n\}$ is  the usual
integration domain and
\begin{equation}
\label{eq:polar_mom} \cF_n(y_i,\zeta_i,p_i)=\exp \sum_{i\neq
j}\left( \half \frac{\zeta_i \cdot \zeta_j}{(y_i-y_j)^2} - \sa \frac{
p_i \cdot \zeta_j}{y_i-y_j} \right) \ ,
\end{equation}
is interpreted with the prescription described above. Note that in this form the $SL(2,\bC)$ symmetry has not been fixed yet. The multi-linear part of $\cF_n$ reads (see \cite{Ademollo:1975pf})
\begin{multline}
\label{eq:F_tensor_structure} \cF_n(y_i,\gz_i) |_{\textrm{multi-linear}}=\sum_{\{i_s\} \in P}
\Bigg\{ \sum_{m=1}^{[n/2]} \Bigg[ \frac{1}{(2m)!!}\frac{1}{(n-2m)!}
\sum_{r=1}^{m} \frac{\gz_{i_{2r-1}} \cdot
\gz_{i_{2r}}}{(y_{i_{2r-1}}-y_{i_{2r}})^2} \times \\ \times
\prod_{s=2m+1}^n \, \left(\sa \sum_{l=1}^n \frac{\gz_{i_s} \cdot
p_l}{y_{i_s}-y_l} \right)\Bigg]\Bigg\} \ ,
\end{multline}
where the global sum is over all sets $\{i_s\}$ of permutations of
indices $\{i_1,i_2,\dots,i_n\}$.

There are several constraints on the kinematical factor arising from (\ref{eq:polar_mom}), mainly coming from fixing the
remnant $SL(2,\bC)$ symmetry. The most useful choice for our purposes is to set $y_1=\infty$, $y_2=1$ and $y_n=0$ which causes the gauge group volume factor in (\ref{eq:boson_vector})
to diverge\footnote{In the Koba-Nielsen amplitude this divergence is canceled
by the term $\prod_i(y_1-y_j)^{2\al p_i \cdot p_j}\simeq y_1^{-2} + {\cal{O}} ( y_1^{-1})$ through
 momentum conservation and the fact that the tachyon mass is
$-\frac{1}{\al}$. This does however not apply in the present case
since the mass-shell condition is  $p_i^2=0$.}
 as $y_1^2$. Any term of the form
\begin{equation}
\label{expexp}
 \sim \left\{ \frac{\gz_i \cdot \gz_j}{(y_i - y_j)^2}\right\}_{(m)}  \left\{ \frac{\gz_h \cdot p_k}{y_h-y_k} \right\}_{(n-2m-1)}
 \frac{\gz_1 \cdot p_l}{y_1-y_l} \ ,
\end{equation}
is then potentially dangerous since it does not cancel the $SL(2,\bC)$
volume completely, giving rise to an integrand that diverges as
$y_1$. However, momentum conservation guarantees that these terms
cancel out in the end, since summing them up for all the different
permutations of external momenta multiplying $\gz_1$ and series
expanding around $y_1 = \infty$ to first order gives for the last factor in  \eqref{expexp}
\begin{equation}
\frac{\gz_1}{y_1} \cdot \sum_l p_l \left(1+\frac{y_l}{y_1}\right) =  \frac{\gz_1 \cdot p_1}{y_1} + \cO\left(\ov{y_1^2}\right) = \cO\left(\ov{y_1^2}\right) \ ,
\end{equation}

To isolate the leading $y_1$ dependence of the integrand, note that the function $\cF_n$ can be expanded in terms of large $y_1$ as
\begin{align}
\cF_n(y_i,\zeta_i,p_i) & = \left(\cF_n \setminus \{1\}\right) \left(\exp \sum_{k=2}^{n} \left( \frac{\zeta_1 \cdot \zeta_k}{(y_1-y_k)^2} - \sa \frac{p_1 \cdot \zeta_k}{y_1-y_k} + \sa \frac{p_k \cdot \zeta_1}{y_1-y_k} \right) \right) \\
& =   \left(\cF_n \setminus \{1\}\right) \left(\exp \sa \sum_{k=2}^{n} \left(\frac{p_1 \cdot \zeta_k}{y_k-y_1} + \left(\frac{ y_k p_k \cdot \zeta_1 + \zeta_1 \cdot \zeta_k}{y_1^2} + \mathcal{O}(\frac{1}{y_1^3})\right) \right) \right)  \ ,
\end{align}
where $ \left(\cF_n \setminus \{1\}\right)$ is $\cF_n$ with all dependence on $y_1$ dropped. Note the written part of $\cF$ above also contains the only dependence on $\zeta_1$. Isolating the linear term in $\zeta_1$ and taking the large $y_1$ limit therefore gives
\begin{multline}
A_n = \int_{\Omega} \prod_{i=3}^{n-1}dy_i \,\,\times \left(\prod_{1<i<j\leq n} (y_i-y_j)^{2 \al p_i \cdot p_j} \right) \left[ \zeta_1 \cdot \zeta_n \left(\cF_n \setminus \{1, \zeta_n\}\right) + \phantom{\sum_{k=2}^{n-1}} \right. \nonumber \\
+ \left. \sum_{k=2}^{n-1} \sa y_k p_k \cdot \zeta_1 \left(\cF_n \setminus \{1\}\right)+ \zeta_1 \cdot \zeta_k \left(\cF_n \setminus \{1, \zeta_k\}\right) \right]\ , \label{eq:boson_vectoramp2}
\end{multline}
where $\left(\cF_n \setminus \{1, \zeta_k\}\right)$ is shorthand for the factor $\cF_n$ of eq.~\eqref{eq:polar_mom} with all dependence on the external polarization $\zeta_k$  dropped, as well as the dependence on $y_1$. Note there is a one term difference between these two operations.

With the explicit shift \eqref{shift_rules} the effect of the BCFW shift on the amplitude can be examined. The shift singles out two particles, $1$ and $n$. In the change of variables the leading behavior of terms of the form $(y_i-y_j)\sim \frac{1}{z}$, while $(y_j) \sim 1$. The latter type of term comes from instances where `$y_n$' would appear, so it is advantageous to isolate these terms. In the just introduced shorthand this leads to
\begin{align}
A_n = \int_{\Omega} \prod_{i=3}^{n-1}dy_i \,\,\times & \left(\prod_{1<i<j\leq n} (y_i-y_j)^{2 \al p_i \cdot p_j} \right) F  \ ,
\end{align}
with
\begin{multline}
F = \zeta_1 \cdot \zeta_n \left(\exp \left( \sum_{j=2}^{n-1} \frac{p_n \cdot \zeta_j}{y_j} \right) \right) \left(\cF_n \setminus \{1, n\}\right)  + \left(\sum_{k=2}^{n-1} \sa y_k p_k \cdot \zeta_1 \right) \times\\ \times\left(\sum_{j=2}^{n-1} \frac{\zeta_n \cdot \zeta_j}{(y_j)^2} - \sa \frac{p_j \cdot \zeta_n}{y_j}  \right)  \left(\exp \left( \sa \sum_{l=2}^{n-1} \frac{p_n \cdot \zeta_l}{y_l} \right) \right) \left(\cF_n \setminus \{1,n\}\right)
 +  \\
 +   \sum_{j,\,k=2}^{n-1} \zeta_1 \cdot \zeta_k \left(\frac{\zeta_n \cdot \zeta_j}{(y_j)^2} - \sa \frac{p_j \cdot \zeta_n}{y_k}  \right) \left(\exp \left( \sa \sum_{j=2}^{n-1} \frac{p_n \cdot \zeta_j}{y_j} \right) \right) \left(\cF_n \setminus \{1,n, \zeta_k\}\right) \ ,
\end{multline}
where again it is understood that the multi-linear term is extracted. Note that in the above all $z$-dependent external polarizations and momenta are written explicitly. Furthermore, from the combinatorics it is easy to count $(y_k)$ versus $(y_i-y_j)$ terms which as mentioned before in the change of variables end up with different z-dependence. The leading behavior in large $z$ comes from terms where the maximum of polarizations are accompanied by $\frac{1}{(y_i-y_j)}$. These arise from $\left(\cF_n \setminus \{1, n\}\right)$. It will be advantageous to choose the gauge
\begin{equation}
q \cdot \zeta_k =0 \qquad \forall \, k \neq 1,n \qquad \textrm{(gauge choice)}  \ ,
\end{equation}
which makes $(p_n \cdot \zeta_j)$ independent of $z$ under the shift. Further note that
\begin{equation}
\sum_{j,\,k=2}^{n-1} ( y_j p^{\mu}_j ) (\frac{p^{\nu}_k}{y_k} ) = \sum_{j,\,k=2}^{n-1} ( p^{\mu}_j ) (p^{\nu}_k ) +  \frac{1}{z} B^1_{\mu\nu}  + \mathcal{O}\left(\frac{1}{z} \right)^2  \ ,
\end{equation}
for some antisymmetric matrix $B^1$.

The $z$-dependence which arises from the 'tachyonic' part of the integrand and measure can be calculated as before. Leaving for a moment the effect of the shifts of the polarization vectors $\zeta_1, \zeta_n$ the large $z$-dependence can be written as
\boxit{\begin{multline} \label{eq:Vleadibosstr}
A_n \sim \left(\frac{1}{z}\right)^{\al(p_1 + p_2)^2} \!\!\! \hat{\gz}_1^{\mu}  \left[z \left(g_{\mu\nu} + B^3_{\mu\nu}\right) h_1\left(\ov{z}\right)+ \left( B^1_{\mu\nu} + B^2_{\mu\nu} \right) h_2\left(\ov{z}\right)  + \mathcal{O}\left(\frac{1}{z} \right)\right] \hat{\gz_n}^{\nu}  \ ,
\end{multline}}
for the gluon amplitude in the bosonic string. Here the hatted quantities have been shifted and $h_i$ are as before polynomial functions of $\frac{1}{z}$ with a non-trivial constant term. Furthermore, $B^2$ is the anti-symmetric matrix
\begin{equation}
B^2_{\mu\nu} =\sa \sum_{j,\,k=2}^{n-1}\left( (\zeta_j)_{\nu} (p_k)_{\mu}  - (\zeta_j)_{\mu} (p_k)_{\nu}\right)  \ ,
\end{equation}
and $B^3$ is the symmetric matrix
\begin{equation}
B^3_{\mu\nu} = - 2 \al \sum_{j,\,k=2}^{n-1}\left( (p_j)_{\mu} (p_k)_{\nu}\right) = - 2 \al (p_1 + p_n)_{\mu} (p_1 + p_n)_{\nu}  \ ,
\end{equation}
which is the main difference compared to the field theory answer in eq.~\eqref{eq:largezfieldt}.

Structurally the same analysis as for Yang-Mills theory (which followed from eq.~\eqref{eq:largezfieldt}) can be applied to the bosonic string gluon amplitude, yielding table \ref{tab:kinconbos}. A cross-check on this table from the explicit expression of the integrated four-gluon amplitude is given in Appendix \ref{fourgluons}. Compared to the field theory limit the difference of the $++$ and $--$ shifts stands out, which originates in the matrix $B^3$. The large $z$ behavior for these is proportional to
\begin{equation}
A(++) \sim \al \left(\frac{1}{z}\right)^{\al(p_1+p_n)^2 -1}  \ .
\end{equation}
The fact that the leading pole is proportional to $\al$ for the $++$ shift suggests that it arises in the field theory from a shift involving the $\al F^3$ term in the bosonic string theory effective action\footnote{We have been informed that this behavior for the $\al F^3$ term can be verified in a complete field theory analysis \cite{henriette}.}. In four dimensions this term gives rise to all-plus and all-minus three point scattering amplitudes which are allowed in this non-supersymmetric theory. In the strict field theory limit the term $B_3$, which is linear in $\al$, can be dropped in equation \eqref{eq:Vleadibosstr} which reduces the analysis of the polarization dependence of the shift exactly to that in \cite{ArkaniHamed:2008yf}.

\begin{table}
\begin{center}
\begin{tabular}{c|c c c}
$\gz_1 \;\backslash \;\gz_n $ &  $-$  &  $+$  & T \\
\hline
$-$                           & $ -1$ & $ +1$ & $ +1$ \\
$+$                           & $ -3$ & $ -1$ & $ -1$ \\
T                             & $ -1$ & $ +1$ & $ -1$ \\
T2                            & $ -1$ & $ +1$ & $  0$ \\
\end{tabular}
\caption{The leading power in $z^{-\al(p_1+p_n)^2-\kappa}$ for large $z$ limit of the adjacent shift of an all gluon amplitude in the \textbf{bosonic string} for all possible polarizations.
  \label{tab:kinconbos}}
\end{center}
\end{table}

\subsubsection{Gluon amplitudes in the superstring}
Consider the supersymmetric, color-ordered amplitude for $n$ gluons
\cite{Green:1987sp}
\begin{multline}
\label{eq:supstr_vector} A_n = (y_A - y_B)(y_B - y_C)(y_A - y_C)
\int_{\Omega} \prod_{i=1}^{n}dy_i d\gt_i d\eta_i \delta(y_A-y_A^0)
\delta(y_B-y_B^0)\delta(y_C-y_C^0)\\ \times \prod_{1 \leq i<j \leq n}
(y_i-y_j-\gt_i \gt_j)^{2 \al p_i \cdot p_j}
\cF_n(y_i,\gt_i,\eta_i,\zeta_i,p_i) \ ,
\end{multline}
where $\cF(y_i,\gt_i,p_i,\zeta_i)$ is the supersymmetric counterpart
of the multi-linear polarization factor in (\ref{eq:polar_mom}) \be
\cF_n(y_i,\gt_i,\eta_i,\zeta_i,p_i) =
\text{Exp}\left\{\sum_{i\neq
j}\frac{\eta_j(\gt_i-\gt_j)p_i\cdot\gz_j \sa - \half \eta_i \eta_j
\gz_i\cdot\gz_j}{y_i - y_j - \gt_i \gt_j} \right\} \ ,
\ee
written in a manifestly supersymmetric form as a Grassmann integral. For the calculation in this subsection and in Appendix \ref{app:antisymsuperstring} we will set $\al=\frac{1}{2}$ to de-clutter the notation, restoring $\al$ dependence by dimensional analysis in the end.

As the calculation will turn out to be much more involved then the bosonic string case, let us present the line of the argument first. Again the choice $y_1=\infty$, $y_2=1$ and $y_n=0$ will be employed, which again causes a $y_1^2$ divergence in the $SL(2,\bC)$ volume factor. The $y_1^2$ terms in the integrand can be isolated most easily by first integrating out $\eta_1$. The simplification brought about by the limit will enable us to integrate out $\theta_1$ rather easily. After this the integrand is in a shape which can be evaluated in the large $z$ limit by careful consideration of all the different contributions.

Integrating out $\eta_1$ and isolating the $y_1$ dependent pieces yields
\begin{multline}
A_n = \int_{\tilde{\Omega}} \left( \prod_{1<i<j\leq n}(y_i-y_j-\gt_i \gt_j)^{p_i \cdot p_j} \right)
\left(\cF_n \setminus \{ 1 \}\right) \left[ \left( \prod_{1<j \leq n}(y_1-y_j-\gt_1 \gt_j)^{ p_1 \cdot p_j} \right)  \right. \\
\left. \left(e^{\sum_{j=2}^{n} \frac{\eta_j(\gt_1-\gt_j) (p_1\cdot\gz_j) }{y_1 - y_j}} \right) \left( \sum_{i=2}^n \frac{(\gt_i-\gt_1)(p_i\cdot\gz_1) + \eta_i (\gz_i\cdot\gz_1)}{y_i - y_1 - \gt_i \cdot \gt_1} \right)    \right]  \ ,
\end{multline}
where $\widetilde{\Omega}$ contains all the intricacies of the integration measure. The three different terms after '$\left(\cF_n \setminus \{ 1 \}\right)$' in the above equation contain all dependence on $y_1$ and $\theta_1$ of the integrand and can each be expanded in terms of $\frac{1}{y_1}$ with the result
\begin{align}
\textrm{I} & = 1 - \frac{1}{y_1} \left( \sum_{k=2}^n  (p_1 \cdot p_k) (y_k+\gt_1 \gt_k)  \right) + \mathcal{O}\left(\left(\frac{1}{y_1}\right)^2\right) \\
\textrm{II} & = 1 + \frac{1}{y_1} \left( \sum_{j=2}^{n} \eta_j(\gt_1-\gt_j) (p_1\cdot\gz_j) \right) + \mathcal{O}\left(\left(\frac{1}{y_1}\right)^2\right)\\
\textrm{III} & = \frac{1}{y_1} \left(\sum_{j=2}^n (\theta_1 - \theta_j) (\gz_1 \cdot p_j)  - \eta_j (\gz_1 \cdot \gz_j) \right) +  \nonumber \\
& \qquad \qquad \qquad \frac{1}{y_1^2} \left(\sum_{j=2}^n y_j (\theta_1 - \theta_j) (\gz_1 \cdot p_j) - (y_j + \theta_1 \theta_j) \eta_j (\gz_1 \cdot \gz_j)\right) + \mathcal{O}\left(\left(\frac{1}{y_1}\right)^3\right)  \ .
\end{align}
The leading term in the $y_1 \rightarrow \infty$ limit in the integrand is proportional to $\frac{1}{y_1}$ and can be simplified to
\begin{equation}
\frac{1}{y_1} \left(\sum_{j=2}^n (\theta_1 - \theta_j) (\gz_1 \cdot p_j)  - \eta_j (\gz_1 \cdot \gz_j) \right) =  - \frac{1}{y_1} \left(\sum_{j=2}^n ( \theta_j) (\gz_1 \cdot p_j) +  \eta_j (\gz_1 \cdot \gz_j) \right)  \ ,
\end{equation}
by momentum conservation and transversality. However, this term does not contain any dependence on $\theta_1$. Since the integrand does not contain any other $\theta_1$ at this order, the potentially disastrous $\frac{1}{y_1}$ term vanishes by fermionic integration over this variable. The term in the integrand proportional to $\left(\frac{1}{y_1}\right)^2$ also simplifies quite considerably using the $\theta_1$ integration. The end result of this integration and the $y_1 \rightarrow \infty$ limit reads
\begin{multline}\label{eq:isolsusyamp1}
A_n = \int_{\tilde{\Omega}} \left( \prod_{1<i<j\leq n}(y_i-y_j-\gt_i \gt_j)^{ p_i \cdot p_j} \right)
\left(\cF_n \setminus \{ 1 \}\right)  \left(\sum_{j=2}^n\left( y_j (\gz_1 \cdot p_j) - \theta_j \eta_j (\gz_1 \cdot \gz_j)\right) \right. \\
\left. - \sum_{j,k=2}^n \left( \theta_j (\gz_1 \cdot p_j)  + \eta_j (\gz_1 \cdot \gz_j) \right) \left(  \eta_k (p_1\cdot\gz_k) + (p_1 \cdot p_k) (\gt_k) \right)\right)  \ .
\end{multline}
The dependence on $y_n$ can be isolated from,
\begin{multline}\label{eq:isolsusyamp2}
\left(\cF_n \setminus \{ 1 \}\right) = \left(\cF_n \setminus \{ 1,2 \}\right) \left[ \left( \prod_{1<j < n}(y_j-y_n-\gt_j \gt_n)^{ p_j \cdot p_n} \right)  \right. \\
\left. e^{\sum_{j=2}^{n-1} \frac{\eta_j(\gt_n-\gt_j) (p_n\cdot\gz_j) }{y_j - y_n}} \left( 1+ \sum_{i=2}^{n-1} \frac{\eta_n(\gt_i-\gt_n)(p_i\cdot\gz_n) - \eta_i \eta_n (\gz_n\cdot\gz_i)}{y_i - y_n - \gt_i \cdot \gt_n} \right)    \right]  \ .
\end{multline}

The resulting expressions appear hopelessly complicated. However, in the large $z$ limit much of the structure turns out to be trivial. This is mainly due to the same observation as in the bosonic string case:
\begin{equation}
y_j \sim 1+ \mathcal{O}\left(\frac{1}{z}\right) \qquad \textrm{so} \qquad (y_i-y_j) \sim \mathcal{O}\left(\frac{1}{z}\right) \qquad \forall i,j \neq n  \ .
\end{equation}
In the following the same gauge choice will be employed as above, $\gz_i \cdot q =0$ $\forall i \neq 1,n$. It will be convenient to split the analysis into three distinct parts: the tachyonic integral,
\begin{equation}\label{eq:suptachpart}
\int_{\tilde{\Omega}} \left( \prod_{1<i<j\leq n}(y_i-y_j)^{ p_i \cdot p_j} \right) \sim \left(\frac{1}{z}\right)^{ (p_1 \cdot p_n)} \left( \left(\frac{1}{z}\right)^{n - 3} + \mathcal{O}\left(\frac{1}{z}\right)^{n - 2} \right)  \ ,
\end{equation}
terms which contain all dependence on polarization vectors which are not $\gz_1$ or $\gz_n$,
\begin{equation}\label{eq:symbdomfermint}
\left( \prod_{1<i<j<n}\left(1 -  (p_i \cdot p_j) \frac{\theta_i \theta_j}{y_i-y_j}\right)^{p_i \cdot p_j} \right) \left(\cF_n \setminus \{ 1,2 \}\right)  \ ,
\end{equation}
where an extra factor has been included for convenience and the remaining terms,
\begin{multline}\label{eq:isolsusyamplead}
 \left[\sum_{j=2}^n\left( y_j (\gz_1 \cdot p_j) - \theta_j \eta_j (\gz_1 \cdot \gz_j)\right) \right. \\
\left. - \sum_{j,k=2}^n \left( \theta_j (\gz_1 \cdot p_j)  + \eta_j (\gz_1 \cdot \gz_j) \right) \left(  \eta_k (p_1\cdot\gz_k) + (p_1 \cdot p_k) (\gt_k) \right)\right] \\
\left[ \left( \prod_{1<j < n}\left(1- \frac{\gt_j \gt_n}{y_j-y_n}\right)^{p_j \cdot p_n} \right)  \right. \\
\left. e^{\sum_{j=2}^{n-1} \frac{\eta_j(\gt_n-\gt_j) (p_n\cdot\gz_j) }{y_j - y_n}} \left( 1+ \sum_{i=2}^{n-1} \frac{\eta_n(\gt_i-\gt_n)(p_i\cdot\gz_n) - \eta_i \eta_n (\gz_n\cdot\gz_i)}{y_i - y_n - \gt_i \cdot \gt_n} \right)    \right]  \ .
\end{multline}
These remaining terms contain all the dependence on the shifted polarizations.

The point of this split is that in the large $z$ limit the expression in eq.~\eqref{eq:symbdomfermint} displays a neat stratification in terms of the number of fermionic variables. Symbolically, this reads
\begin{equation}
\sim \sum_{j=2}^n z^{n-j} (\eta)^{n-j} \left( h_1\left(\frac{1}{z}\right) (\theta)^{n-j} + \frac{1}{z} h_2\left(\frac{1}{z}\right) (\theta)^{n-j-2} + \mathcal{O}\left(\frac{1}{z}\right) \right.  \ ,
\end{equation}
where $h$ are as before certain polynomials with non-trivial constant term. The rest term is of the form $\sim \textrm{constant} + \frac{1}{z}$. Using the fermionic integration now allows one to isolate the leading and sub-leading terms in the $z \rightarrow \infty$ limit. The leading term arises from those terms in \eqref{eq:symbdomfermint} with the maximum amount of $\theta$ and $\eta$ integrations. Hence in the rest term, eq.~\eqref{eq:isolsusyamplead}, there cannot be any fermionic variables other than $\theta_n$ and $\eta_n$ left. This yields
\begin{align}
(A(z))_{\textrm{leading}}\sim & (1+ p_1\cdot p_n) \left(\gz_1 \cdot \gz_n\right) + \left[ (p^{\mu}_n) (p^{\nu}_1) - \sum_{j,k=2}^{n-1} \left( y_j p^{\mu}_j \right) \left(\frac{p^{\nu}_j}{y_k} \right) \right] \nonumber \\
\sim & (1+ p_1\cdot p_n) \left(\gz_1 \cdot \gz_n\right) + \frac{1}{z} \gz_1^{\mu} B^1_{\mu\nu} \gz^{\nu}_n  + \mathcal{O}\left(\frac{1}{z} \right)^2  \ ,
\end{align}
for some anti-symmetric matrix $B^1$. The sub-leading terms from eq.~\eqref{eq:isolsusyamplead} can also be calculated. Due to its length this calculation will be deferred to appendix \ref{app:antisymsuperstring}. With this result in hand the large $z$ behavior for the gluonic amplitude in the superstring can be derived from
\boxit{\begin{multline}\label{eq:gluonssuperstringleadingz}
A_n \sim \left(\frac{1}{z}\right)^{2 \al (p_1 \cdot p_n)}  \hat{\gz}_1^{\mu} \left( z \, h_1 (1+2 \al p_1 \cdot p_n) g_{\mu\nu} +  \left(h_2 B^1_{\mu\nu} + h_3 B^2_{\mu\nu} \right)+ \mathcal{O}\left(\frac{1}{z} \right) \right) \hat{\gz}_n^{\nu}  \ .
\end{multline}}
In this formula dependence on $\al$ has been restored through dimensional analysis. The function $h$ are as before polynomial functions of $\left(\ov{z}\right)$ with non-trivial constant term. Importantly, the matrix $B^2_{\mu\nu}$ is antisymmetric as is shown in appendix \ref{app:antisymsuperstring}. By the same analogy to \cite{ArkaniHamed:2008yf} as noted above this leads immediately to Table \ref{tab:kinconsup} for the large $z$-behavior of the superstring gluon amplitude. As an explicit example the analysis for the four-gluon amplitude in type I superstring theory  is carried out in Appendix \ref{fourgluons} which is a $D$-dimensional consistency check of the calculation in \cite{Boels:2008fc}.

\begin{table}
\begin{center}
\begin{tabular}{c|c c c}
$\gz_1 \;\backslash \;\gz_n $ &  $-$  &  $+$  & T \\
\hline
$-$                           & $ +1$ & $ +1$ & $ +1$ \\
$+$                           & $ -3$ & $ +1$ & $ -1$ \\
T                             & $ -1$ & $ +1$ & $ -1$ \\
T2                            & $ -1$ & $ +1$ & $  0$ \\
\end{tabular}
\caption{The leading power in $z^{-\al(p_1+p_n)^2-\kappa}$ for the large $z$ limit of the adjacent shift of an all gluon amplitude in the \textbf{superstring} for all possible polarizations.
  \label{tab:kinconsup}}
\end{center}
\end{table}

\subsubsection*{Application to known MHV amplitudes up to 6 points}
As a check on the general results, it is also interesting to consider the specific case of four-dimensional MHV amplitudes. For the four-point case this was done in \cite{Boels:2008fc}. Here, we consider explicitly the five- and six-point case.

Using the results of Stieberger and Taylor and the analysis of
generic integrals appearing in $n$-point functions given above, it
is straightforward to re-derive the large shift behavior for MHV amplitudes in
four dimensions. For five-point MHV amplitudes we find that in the
notation of \cite{Stieberger:2006te} the contributing hypergeometric
functions behave under shifts of particles $1$ and $n$ as
\begin{equation}
f_1 \sim \left( \frac{1}{z} \right)^{\al (p_1 + p_n)^2 + 2} \quad
\quad f_2 = \left( \frac{1}{z} \right)^{\al (p_1 + p_n)^2 +1} \ .
\end{equation}
Combined with the kinematic pre-factors, this exactly reproduces the results found for the four-point function in the supersymmetric case \cite{Boels:2008fc}.

For the six-point MHV amplitudes in four dimensions there is a six element basis of hypergeometric functions. Several bases are known in the literature, as for instance given in \cite{Oprisa:2005wu} or \cite{Stieberger:2006te}. For our purposes it is most useful to use the one in \cite{Stieberger:2007jv}. The basis elements shift as
\begin{equation}
\begin{array}{lcr}
K_1 \sim \left( \frac{1}{z} \right)^{\al (p_1 + p_n)^2 + 1} & \quad\quad & K_4 \sim \left( \frac{1}{z} \right)^{\al (p_1 + p_n)^2 + 1 } \\
K_2 \sim \left( \frac{1}{z} \right)^{\al (p_1 + p_n)^2 + 1} & \quad\quad & K_5 \sim \left( \frac{1}{z} \right)^{\al (p_1 + p_n)^2 + 2}\\
K_3 \sim \left( \frac{1}{z} \right)^{\al (p_1 + p_n)^2 + 2} & \quad\quad & K_6 \sim \left( \frac{1}{z} \right)^{\al (p_1 + p_n)^2
+ 1}
\end{array}  \ ,
\end{equation}
for the same shift as considered above. The coefficients in the expression for the amplitude scale individually like $z^1$, seemingly leading to a different recursion condition. However, the leading poles of the summed expression cancel pairwise between different terms leading again to the constraints of table \ref{tab:kinconsup}. The expression in \cite{Stieberger:2007jv} is the helicity configuration $\braket{--++++}$ only, but the others are related by the supersymmetric Ward identity just as in field theory. In all, this yields the same picture as in the five particle case. The seven-point MHV amplitude, given in Ref.~\cite{Stieberger:2007jv}, could be analyzed in a similar way at least in principle.

In general the derived large shift behavior of the amplitudes implies interesting cross-relations between basis coefficients in the `expand in basis functions' approach as developed in \cite{Stieberger:2007jv},\cite{Stieberger:2006te}, \cite{Stieberger:2007am}.

\subsubsection*{General MHV amplitudes in the superstring}

In the appendix of Ref.~\cite{Berkovits:2008ic}, a superspace formula was written which is conjectured to be the full $d=4$ MHV amplitude in the superstring. Again, the integrals are similar to the bosonic case. In view of the results above, it would be interesting to examine this conjectured formula in more detail. In particular, if it can be proven that the formula
\begin{itemize}
\item shares the same spectrum as the ordinary superstring amplitude
\item shares the same three point couplings as the ordinary superstring amplitude
\item obeys the same recursion relations
\end{itemize}
as the ordinary superstring, then the conjectured formula is indeed the complete amplitude. For the first point to hold, all factorizations of the four, five and six point amplitude in one channel should be studied. The second point can be checked by studying the factorization of the six point function in three channels. The last point can be checked by studying the large $z$-behavior along the lines given above.

\subsection{All open string tree amplitudes in a flat background}
In \cite{Boels:2008fc} it was pointed out that the analysis for the four-point bosonic open string tachyon amplitude generalizes to all
four-point string amplitudes in any open string theory in a flat background. This follows simply because any such amplitude will
involve sums over integrals of the Veneziano type. By a similar reasoning, the result above for shifts of the $n$-point bosonic string tachyon amplitude extends to any open string theory amplitude in a flat background. These amplitudes therefore \emph{all} obey
on-shell recursion relations. The remaining question to be answered is under which constraints this holds. It is clear that generically for a shift of legs $1$ and $n$ any open string amplitude will scale as
\begin{equation}
A_n(z) \rightarrow \left(\frac{1}{z}\right)^{\al (p_1 + p_n)^2 - \kappa} G  \ ,
\end{equation}
for some $z$-independent function $G$ and some integer $\kappa$. This reduces the analysis of on-shell recursion in open string theory to the determination of an integer.

In fact, based on the above examples we expect that the integer $\kappa$ is `universal': for a given amplitude it should only depend on the quantum numbers of the two particles being shifted. In other words, we expect that this integer does not depend on the particle content of the rest of the amplitude. In the CFT argument of section \ref{sec:CFTderivation} this universality property will be manifest.

There are a host of further explicit examples which one could check, including for example string amplitudes with massive particles, RR fields\footnote{See e.g. Refs.~\cite{Liu:1987tb,Xiao:2005yn} for results on three- and four-point superstring amplitudes involving massive string states at the first excited level and \cite{Hatefi:2010ik} (and references therein) for examples involving a RR field.} and superstring amplitudes involving fermions using the same technique as above.  As the kinematic constraints are expected to be universal as explained above, the four-point function contains valuable information on explicit constraints. A more direct way could be to apply the CFT arguments of Section \ref{sec:CFTderivation} to these cases. Based on the relation to the field theory limit of the gluon shifts in the superstring, it is easy to formulate an expectation for the shifts of the fermions. Finally we note that the supershift of \cite{ArkaniHamed:2008gz} for massless particles also follows from the above as the behavior under this shift can be derived from the ordinary BCFW shift. The extension of this supershift to massive states is highly interesting.

\section{Recursion relations for the disc: non-adjacent shifts}
\label{sec:nonadjshift}
To analyze non-adjacent shifts certain relations between different color orders first discussed in \cite{Plahte:1970wy} (see \cite{BjerrumBohr:2009rd} and \cite{Stieberger:2009hq} for a more modern perspective) will be used to relate non-adjacent shifts to adjacent ones. These relations will be referred to as monodromy relations for reasons which will become obvious in the course of the discussion.

\subsection{Veneziano re-revisited}
It is instructive to start the discussion by visiting the Veneziano amplitude yet again. For four particles the monodromy relations read \cite{Plahte:1970wy}
\begin{equation}\label{eq:mon4pt1}
A(1234) + e^{ \ii 2 \pi  \al p_1 p_2} A(2134) + e^{ \ii  2 \pi \al
p_1 (p_2+ p_3)} A(2314)  = 0 \quad \Im(y)>0 \ ,
\end{equation}
and
\begin{equation}\label{eq:mon4pt2}
A(1234) + e^{-\ii  \pi \al 2 p_1 p_2} A(2134) + e^{-\ii  2 \pi \al
p_1 (p_2+ p_3)} A(2314)  = 0 \quad \Im(y)< 0 \ ,
\end{equation}
where $y$ is the integration variable in \eqref{eq:venezianoampintegral}.
These can be derived from the integral representation
\eqref{eq:venezianoampintegral} by extending the boundaries of the
integral to (above or below) the complete real $y$-axis and closing
the contour. The two half-infinite integrations along the real axis
can be related to amplitudes by gauge fixing vertex operators in a
different order, picking up phase factors from the branches of the
logarithms. For our purposes, these relations allow one to relate a
non-adjacent shift to an adjacent one.
To do this properly there would appear to be a sign problem though: naively when shifting either the relation \eqref{eq:mon4pt1} or the relation \eqref{eq:mon4pt1} will have divergent behavior. Obviously, they can't both hold. Let us therefore derive the conditions under which the relations hold.

One source of potential problems is the boundary integral on the half-arc at infinity. For this to be shown to vanish,
the behavior of the integrand needs to be studied. In the limit $|y|
\rightarrow \infty$ this reads
\begin{equation}\label{eq:exposeorderoflimits}
y^{\al s - 2} (1-y)^{\al t  - 2} \sim e^{- \al u \log(|y|) + i (\al
s - 2) \arg (y) + i (\al t - 2) \arg (1-y) } \ .
\end{equation}
Hence under the reality condition
\begin{equation}
\Re(\al u) > 1 \ ,
\end{equation}
the integrand on the half-arc vanishes in the limit $|y|\rightarrow
\infty$. However, this ignores an important subtlety: the relations
\eqref{eq:mon4pt1} and \eqref{eq:mon4pt2} will be employed to study
the non-adjacent shift of particles $1$ and $3$ in the limit $z'
\rightarrow \infty$ as in eq.~\eqref{eq:shiftmandvar}. Hence there exists the possibility of an order
of limits problem. Under the non-adjacent shift one has
\begin{equation}
\al s \rightarrow \al s + z' \ , \quad\quad \al t \rightarrow \al t - z'
\ ,
\end{equation}
so that the integrand \eqref{eq:exposeorderoflimits} picks up a
phase factor on a half-arc of radius $\epsilon$
\begin{equation}
\sim e^{\pm i z' (\pi  + \mathcal{O}(\epsilon))} \ .
\end{equation}
Here the sign depends on whether the contour is closed above $(+)$
or below $(-)$ the real axis. Therefore either \eqref{eq:mon4pt1} or
\eqref{eq:mon4pt2} is available as an identity valid in the $z'
\rightarrow \infty$ limit, in this case without reality conditions.
It thus follows that both for $\Im(z')>0$ and $\Im(z')<0$ the
integrand for the residue integral at infinity vanishes
exponentially.

In detail, the large-$z$ behavior of the non-adjacent shift reads,
\begin{eqnarray}
A(1234) \sim  e^{ i z'} (z')^{-\al u -1 } \left(G_0+ \mathcal{O}(\frac{1}{z'})\right) \ , & \quad & \Im(z') > 0 \\
A(1234) \sim  e^{- i z'} (z')^{-\al u -1 } \left(\tilde{G}_0 +
\mathcal{O}(\frac{1}{z'})\right)  \ ,  & \quad & \Im(z') < 0  \ ,
\end{eqnarray}
which confirms what was obtained in \cite{Boels:2008fc} by an
argument based on the behavior of the Gamma function. The residue at
infinity therefore vanishes exponentially by a simple extension of
the argument above, after excising the special points on the real
$z'$ axis and application of the same analyticity argument as before.

Actually in addition to the pole at infinity which might be
potential problem for the analysis of the monodromy relations, there
are two poles at both $y=0$ and $y=1$  in eq.~\eqref{eq:venezianoampintegral} which deserve some further
attention. Using a half-arc integral around the potentially
dangerous points yields two further reality conditions,
\begin{equation}
\Re(\al  s) > 1 \ ,
\end{equation}
and
\begin{equation}
\Re(\al t) > 1  \ ,
\end{equation}
again, as long as the arc-integral itself is finite. See below for a physical interpretation of these equations. This completes the analysis of the reality conditions in the four-point case.

\subsection{Amplitude monodromy relations from the CFT}\label{sec:monfromcft}
The monodromy relations can be generalized to all open string amplitudes by direct examination of the Koba-Nielsen type integral representation, at least in principle. As will be shown here they can also be elegantly derived by appealing directly to the CFT origin of open string amplitudes.

The crucial observation is that the open string vertex operators of string theory in a flat background have simple co-cycle factors
\begin{equation}\label{eq:braidingsignfactor}
:\!V_1(z_1)\!:\, :\!V_2(z_2)\!:\, \equiv \,:\!V_2(z_2)\!:\, :\!V_1(z_1)\!:\, e^{ 2 \pi \ii \al
\left(p_1 p_2\right) \epsilon({z_1,z_2})}  \ ,
\end{equation}
where $\epsilon({z_1,z_2}) = \pm (z_1-z_2)/|z_1-z_2|$ (see e.g.
\cite{Green:1987sp}, equation (7.1.49)). This follows simply from
the operator product expansion as the sign arises from a choice of
branch cut of the logarithm in the contraction between string
fields. As shown below, it is this factor which underlies the
monodromy relations.

The color ordered multi-field string amplitude is obtained from CFT
correlation functions by the usual prescription,
\begin{equation}\label{eq:genopenstringamp}
A(1, \ldots, n) = \int' \theta(z_i) \langle 0|:\!V(z_1)\!: \ldots :\!V(z_n)\!: | 0 \rangle  \ ,
\end{equation}
where $V$ are vertex operators of physical states in the string
theory. Furthermore $\int'$ is the conformal invariant integration
over the insertion points including the measure and $\theta(z_i)$
enforces the integration domain to be restricted to the ordering of
operators along the edge of the disc indicated in the correlation
function. By moving $:V(z_1):$ through the vertices and tracking the
signs from eq.~\eqref{eq:braidingsignfactor} it is easy to see
the amplitude is cyclic, but only if the sign choice is universal.

\begin{figure}[!ht]
  \begin{center}
 \includegraphics[scale=0.3]{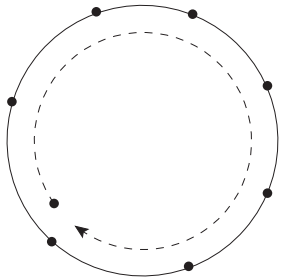}
  \label{fig:contour}
\caption{Choice of contour on the interior of the disc.}
  \end{center}
\end{figure}

A first example of monodromy relations can be obtained directly from
eq.~\eqref{eq:braidingsignfactor} by studying a contour
integration of the CFT correlation function over one of the
insertion points as
\begin{equation}\label{eq:cftmonodromeq}
\oint_{z_1} \langle 0|:\!V(z_1)\!: \ldots :\!V(z_n)\!: | 0 \rangle = 0  \ .
\end{equation}
The contour is located on the interior of the disc (see Figure
\ref{fig:contour}) and since the correlation function is analytic
over the interior the integral vanishes. The other vertex operators
are assumed to be color ordered. It is easy to see that the expression on
the left hand side of this equation is a sum over the integrands of
string amplitudes as in eq.~\eqref{eq:genopenstringamp} -
almost. The difference is that the vertex operator for particle one
needs to appear ordered in the correlation function. Using the
relation \eqref{eq:braidingsignfactor} this can be done
straightforwardly. Hence the equality
\begin{equation}\label{eq:monodromyexampleone}
A(\alpha_1,1,\beta_1 \ldots \beta_k , n) = -  \sum_{\sigma \in
OP\{\alpha_1\} \cup \{\beta\}} e^{\pm (\alpha_1 1)} \left(
\prod_{i=1}^k e^{\pm(\alpha_1 \beta_i)} \right)  A(1, \sigma, n)  \ ,
\end{equation}
is derived as a first monodromy relation between amplitudes. Here
the sum is over all permutations which preserve the order of the
subsets and the phase factor
\begin{equation}\label{eq:phasefac}
e^{\pm (i j)} = \left\{ \begin{array}{ccl} e^{\pm 2 \pi  \ii \al
p_{i} p_{j}} & \quad & \textrm{if} \quad z_j < z_i \\ 0 & \quad &
\textrm{else} \end{array} \right.  \ ,
\end{equation}
depends on the ordering of the particles within the particular amplitude
that this factor multiplies. Note that the entire argument only
depends on the monodromy relation  \eqref{eq:braidingsignfactor},
the topology of the disc and the assumption that there are no poles in the interior.
It therefore holds quite generally for open string amplitudes, up to two subtleties mentioned below.

The sign in the relation derives from the choice of sign in eq.~\eqref{eq:braidingsignfactor}.
This choice however does not affect the calculation of the amplitudes themselves and hence the relation \eqref{eq:monodromyexampleone} holds for both choices of sign. Note that the amplitudes are not considered to be real (as in
\cite{BjerrumBohr:2009rd,Stieberger:2009hq}), which is a special choice only available with a definite reality condition on
the momenta. As the relations will be used to study BCFW shifts in this paper (which are inherently complex), this is an important property.

\subsection*{Two subtleties}
There are two subtleties related to limits within the above argument which will be important in the following. Looking more closely to the process in which the integration contour is taken to the edge of the disc shows
that a careful limit argument is necessary, by taking for instance
half-arcs around the vertex operator insertions. The integration
over these arcs with, say, radius $\epsilon$ can easily be performed
by considering the OPE of the `stationary' and the moving vertex
operator,
\begin{align}
\int_{R^+_{\epsilon}(z_i)} dz_1 \langle :\!V(z_1)\!: :\!V(z_i)\!:   \ldots \rangle & = \int_{R_{\epsilon}(z_i)} \langle  \left(z_1 - z_i \right)^{2\al  p_1 \cdot p_i} :\!V(z_1) V(z_i)\!:   \ldots \rangle \\
& = \int_{R_{\epsilon}(z_i)} dz_1 \langle  \left(z_1 - z_i
\right)^{2\al p_1 \cdot p_i} \sum_{k=0}^{\infty} C_{1ik} (z_1
-z_i)^k :\!V_k\!:   \ldots \rangle \ .
\end{align}
Here $C_{1ik}$ are the CFT structure constants. Note that a similar
computation yields the leading pole behavior for kinematic limits of
particles $1$ and $i$ of the form $\al(p_1 + p_i)^2 \rightarrow m $
for some integer $m$. The `leading singularity' of the integral is
therefore found by considering the coupling of particles $1$ and $i$
to the lowest mass resonance possible in the theory. For bosonic string
tachyons for instance this is the tachyon itself.

Since the above formula isolates the singular behavior in the limit
$\epsilon \rightarrow 0$ it allows the corresponding integral to be
evaluated to give
\begin{equation}
\int_{R^+_{\epsilon}(z_i)} dz_1 \left(z_1 - z_i \right)^{2\al  p_1
\cdot p_i} \rightarrow \epsilon^{-1 - 2\al  p_1 \cdot p_i }
\int_{0}^{\pi} d\theta e^{\ii \theta (2\al  p_1 \cdot p_i)} \ .
\end{equation}
Hence the arc integral vanishes for the open bosonic string tachyon
amplitude in the limit $\epsilon \rightarrow 0$ if
\begin{equation}
\Re (2\al  p_1 \cdot p_i) > 1 \quad \quad \textrm{if $\theta$
integral finite} \ ,
\end{equation}
as long as the angle integral over $\theta$ is finite. In other
words, the monodromy relation \eqref{eq:monodromyexampleone} holds
as long as the momentum invariant $p_1 \cdot p_i$ is away from the
first resonance that the particles one and $i$ couple to.

\subsubsection*{Subtlety 1}
Based on the above it would seem that the monodromy relation is
invalid in the limit where
\begin{equation}
2 \al  p_1 \cdot p_i \rightarrow 1 \ .
\end{equation}
This is exactly the location of the tachyon pole. However, there are
two limits here and it might in principle matter in which order they
are taken. If first $\epsilon \rightarrow 0$, then the tachyon pole
is contained in the string amplitudes since the integration along
the edge of the disc is all the way up to the singular point. If
first $2 \al p_1 \cdot p_i \rightarrow 1$ is taken, then the pole is
in the arc integral and should be accounted for separately. In both
cases, the same result is obtained but form different parts of the
calculation. Hence there is no order of limits problem for these
particular two limits. This is the first of the two subtleties
referred to above.

\subsubsection*{Subtlety 2}
A second more serious subtlety has to do with the actual integral
over the arc. As referred to above, this integral is assumed to be
finite in order to apply the above argument. This assumption can
easily be violated when the momenta are taken to (complex) infinity.
The result is then a $0$ times $\infty$ type ambiguity. For general
BCFW shifts of particle one and $n$ this implies there is only one
choice of sign which is allowed, depending on the direction in the
complex plane the BCFW limit is taken in. Note that the choice of
sign for which the half-arc integrals vanish corresponds to the same
choice of sign in the general relations. Hence in general monodromy
relations will lead to exponentially suppressed amplitudes for
non-adjacent BCFW shifts, without reality conditions. This will be
spelled out below.

\subsection*{General monodromy relations}
Before studying shifts it is interesting to consider generic
applications of the relations. To derive a general result consider
\begin{multline}\label{eq:cftmonodromeq2}
\int' \prod_{i=1}^{k} \oint_{z_{\alpha_i}} H(|z_{\alpha_i} - z_{1}|
> |z_{\alpha_{i+1}} - z_1|) \\ \langle 0|:\!V(z_{\alpha_1})\!: \ldots
:\!V(z_{\alpha_k})\!: :\!V_1\!: :\!V_{\beta_1}\!: \ldots  :\!V_{\beta_s}\!: :\!V(z_n)\!:
| 0 \rangle = 0 \ ,
\end{multline}
and fix the position of operators $1$, $n$ and one of the $\beta$'s.
The Heaviside $H$ function forces an ordering of the $\alpha$
particles (operators at $z_{\alpha_i}$) with respect to the distance
to the first fixed operator. Again, ordering the operators to
reproduce the amplitude integrands gives
\begin{multline}\label{eq:towardbjbdamvhresult}
A(\alpha_1 \ldots \alpha_r 1 \beta_1 \ldots \beta_s n) = \\(-1)^r
\left( \prod_{i} e^{\pm (\alpha_i 1)} \right) \left( \prod_{i>j}
e^{\pm (\alpha_i \alpha_j)} \right) \left( \sum_{\sigma \in
OP\{\alpha\}^T \cup \{\beta\}} \left( \prod_{i=1}^r \prod_{j=1}^{s}
e^{\pm (\alpha_i \beta_j)}\right) A(1 \sigma n)  \right) \ ,
\end{multline}
in terms of the phase factors in eq.~\eqref{eq:phasefac}. These phase factors depend on the ordering of the particles in the term under study of the sum. The conditions under which this is allowed are unchanged with respect to the analysis above, including the remarks about the subtleties. Note that the same sign should be used for all shuffling operations in the co-cycle factor in eq.~\eqref{eq:braidingsignfactor}. This leaves one sign choice in the result \eqref{eq:towardbjbdamvhresult}. These relations show how to express any amplitude into ones with a fixed order of two particles. Using first one sign choice and then the other allows one to find a basis of all amplitudes with three particles in a fixed order. This basis has $(n-3)!$ elements, as shown explicitly in \cite{BjerrumBohr:2009rd}.

\subsubsection*{Derivation by iteration}
The same result can also be derived by induction from \eqref{eq:monodromyexampleone}. The possibility of
doing this is mentioned in \cite{Stieberger:2009hq}. The
base step was proven above. For the induction step assume
\eqref{eq:towardbjbdamvhresult} holds for $(r-1)$ $\alpha$
particles. From \eqref{eq:monodromyexampleone} the following
equation holds by pushing $\alpha_1$ through the amplitude,
\begin{multline}\label{eq:pushingthroughonealpha}
A(\alpha_1 \ldots \alpha_r 1 \beta_1 \ldots \beta_s n) +  e^{\pm
(\alpha_2 \alpha_1)} A(\alpha_2 \alpha_1 \ldots \alpha_r 1 \beta_1
\ldots \beta_s n)  + \ldots = \\ - C(\alpha_1 \ldots \alpha_r 1
\beta_1 \ldots \beta_s n) - e^{\pm (\alpha_2 \alpha_1)} C(\alpha_2
\alpha_1 \ldots \alpha_r 1 \beta_1 \ldots \beta_s n) + \ldots \ ,
\end{multline}
where terms with the same number of particles between $1$ and $n$
have been collected on the left hand side, moving the others to the
right. Here $C({\alpha}, 1,{\beta}, n)$ is defined as the right hand
side of eq.~\eqref{eq:towardbjbdamvhresult}. The terms of the
right hand side are split into different orderings of $\alpha_1$
amongst the ordered other $\alpha$'s. For every ordering on the
right hand side there is a corresponding (inverse) ordering of
$\alpha$'s on the left hand side. Indeed, grouping terms shows
already \eqref{eq:towardbjbdamvhresult}. By re-labeling the $\alpha$
particles cyclically more equations can be derived, leading to a
system of $s$ independent inhomogeneous linear  equations in $s$ unknowns. The unique
solution of this system reproduces \eqref{eq:towardbjbdamvhresult}.
Again, the derivation uses \eqref{eq:monodromyexampleone} for one
choice of sign only.

\subsection*{Extension to fermions}
The above argument can also easily be adapted to study monodromy
relations for amplitudes which involve fermions. From the field
theory point of view examples of these in four dimensions have been
discussed in \cite{Sondergaard:2009za}. For this one needs the
co-cycle factors
\begin{equation}\label{eq:braidingsignfactorfermions}
:\!V^f_1(z_1)\!: :\!V^f_2(z_2)\!: \equiv - :\!V^f_2(z_2)\!: :\!V^f_1(z_1)\!: e^{i 2
\pi \al \left(p_1 p_2\right) \epsilon({z_1,z_2})} \ ,
\end{equation}
with an extra minus sign  from Fermi statistics, and
\begin{equation}\label{eq:braidingsignfactorfermionsII}
:\!V_1(z_1)\!: :\!V^f_2(z_2)\!: \equiv :\!V^f_2(z_2)\!: :\!V_1(z_1)\!: e^{i 2 \pi
\al \left(p_1 p_2\right) \epsilon({z_1,z_2})} \ ,
\end{equation}
where the superscript indicates if a vertex operator creates a
fermionic state or not. The derivation of general relations then
proceeds along similar lines to the above, subject to similar
reality conditions.

\subsection{Non-adjacent shifts from monodromy relations}
The study of monodromy relations in the previous section allows an
elementary derivation of the non-adjacent shift behavior of string
theory amplitudes. For concreteness we first focus on open bosonic
string tachyon amplitudes. The first example follows from studying
eq.~\eqref{eq:monodromyexampleone} which can be used  to shift
particles $1$ and $n$ to yield
\begin{equation}\label{eq:behaviorwithonealpha}
A(\alpha_1 1 \beta_1 \ldots \beta_s n) = e^{\pm \ii  2 z  q
p_{\alpha_1}} z^{2 \al p_1 p_n + 1 } \left( G_0 +  \frac{G_1}{z} +
\mathcal{O}(z) \right) \ ,
\end{equation}
in this expression the $+$ sign should be chosen if $\Im (2z  q
p_{\alpha_1})>0$ and the $-$ sign if $\Im (2z q p_{\alpha_1})< 0$.
Note that the behavior of the adjacent shift is simply repeated -
suppressed by an exponential. In a real sense this equation is an
upper limit: the in principle complicated function $G_0$ could
vanish.

As can be seen from eq.~\eqref{eq:pushingthroughonealpha} in a
case where there are more than $1$ particles between the particles
which are shifted, the analysis becomes more complicated. To get a
firm grasp of what is needed, let us study the case of two $\alpha$
particles. In this case in principle $4$ types of relations can be
derived by pushing either $\alpha_1$ or $\alpha_2$ through, with two
sign choices each. Based on the subtleties mentioned above, only $2$
of these types of relations hold in the large $z$ limit, depending
on the sign of the two quantities,
\begin{equation}
\Im (2 z q p_{\alpha_1}) \quad, \Im (2 z q p_{\alpha_2})  \nonumber \ .
\end{equation}
For a fixed choice of sign of these quantities, the two types of
relations which do hold involve two unknowns,
\begin{equation}
A(\alpha_1 \alpha_2 \hat{1} \beta \hat{n} ) \quad \textrm{and} \quad
A(\alpha_2 \alpha_1 \hat{1} \beta \hat{n} ) \ ,
\end{equation}
since the behavior of amplitudes with only one $\alpha$ is known from eq.~\eqref{eq:behaviorwithonealpha}. However, on inspecting it is seen that there are cases where the two equations obtained are degenerate as the large z-shift of the relations read,
\begin{multline}
A(\alpha_1 \alpha_2 \hat{1} \beta \hat{n} )+ e^{\pm_1 (\alpha_1
\alpha_2)} A(\alpha_2 \alpha_1 \hat{1} \beta \hat{n} ) \\ = -
e^{\pm_1 \ii z 2 q p_{\alpha_1}} e^{\pm_2 \ii z 2 q p_{\alpha_2}}
z^{2 \al p_1 p_n + 1 } \left( G_0 +  \frac{G_1}{z} + \mathcal{O}(z)
\right) \ ,
\end{multline}
\begin{multline}
e^{\pm_2 (\alpha_2 \alpha_1)} A(\alpha_1 \alpha_2 \hat{1} \beta
\hat{n} )+ A(\alpha_2 \alpha_1 \hat{1} \beta \hat{n} ) \\ = -
e^{\pm_1 \ii z 2 q p_{\alpha_1}} e^{\pm_2 \ii z 2 q p_{\alpha_2}}
z^{2 \al p_1 p_n + 1 } \left( G_0 +  \frac{G_1}{z} + \mathcal{O}(z)
\right) \ ,
\end{multline}
where the choice of signs $(\pm)_{1,2}$ in every contour is such that there is exponential suppression. To get around this the same modification of the integrand as in \eqref{eq:towardbjbdamvhresult} will yield a relation where the $\alpha_i$ particles appear ordered. In particular, there is only one term in the relation with all $\alpha_i$ particles between the shifted particles $1$ and $n$, while all the other terms contain adjacent particles $1$ and $n$ only.

This yields for a shift of generically non-adjacent particles $1$ and $j$
\boxit{\begin{equation}\label{eq:gennonadjshift}
A(z) \sim e^{ 2 \al z q \cdot\left(\sum_{i=2}^{j-1} \pm_i p_i \right)} z^{\al (p_1 + p_j)^2} \left( G(z) + \mathcal{O}\left(\frac{1}{z}\right) \right.  \ ,
\end{equation}}
where $G(z)$ starts with the same power of $z$ as the adjacent shift of particles $1$ and $j$ and the signs in the exponential are such that the resulting shift is always exponentially suppressed, regardless of the direction of $z$. This argument holds up to $2 ((j-1)-1)$ points on the circle at infinity. Excising these and inserting our analyticity argument as before finishes the discussion of non-adjacent shifts, up to one subtlety.

This subtlety has to do with the field theory limit: in this limit the field theory is known to behave one power of $z$ better than the adjacent shift. An argument in the string theory for this which will be substantiated more in the next section is as follows. For the case of one $\alpha$ particle (Next-to-Adjacent shifts) eq.~\eqref{eq:gennonadjshift} is special.  From \eqref{eq:behaviorwithonealpha} in the large $z$ limit the non-adjacent shift is related to a sum over terms which look like a monodromy relation again, but now with an `effective' particle made out of the shifted particles, whose momentum is the sum of both. This does not happen for more $\alpha$ particles because of the complicated sign structure. If the leading power can be interpreted as a physical state in the string again, then the leading power of $z$ cancels exactly in this particular case. Below an argument for this from the CFT will be given.

\section{OPE derivation of adjacent shifts}
\label{sec:CFTderivation} The results above for the kinematic constraints for the particle shifts of adjacent gluon legs beg for a more streamlined and above all physical discussion. From the point of view of the string theory, it is natural to suspect that this discussion should be in the language of the underlying CFT. As the analysis for non-adjacent shifts is already in this form, what is needed to complete the CFT picture is the adjacent shift behavior. For four points this is closely related to Regge behavior, which has been analyzed directly from the underlying CFT in \cite{Brower:2006ea} for other purposes. Below the approach of that paper is adapted to the case at hand.

\subsection{Shifting tachyons}
Let us gauge fix operators $2$, $n-1$ and $n$ to $0$, $1$ and $\infty$ respectively for an $n$-particle color-ordered amplitude,
\begin{equation}
A(1 2 \ldots n) \ ,
\end{equation}
and consider the shift,
\begin{equation}
p_1 \rightarrow p_1 + z q \quad , \quad p_2 \rightarrow p_2 - z q \ ,
\end{equation}
with the usual constraint \eqref{eq:constraintnmu}. The shifted particles will be taken to be tachyons, while the rest of the amplitude is left general. From the OPE of the two open string tachyon vertex operators
\begin{equation}\label{eq:afetrtachope}
:\!e^{\ii \hat{p}_1 X(y)}\!: \, :\!e^{\ii \hat{p}_2 X(0)}\!: = \left(y\right)^{2
\al p_1 p_2} :\! e^{\ii \left(p_1 X(y) + p_2 X(0)\right) + \ii z
q^{\mu} \left(X_{\mu}(y) - X_{\mu}(0)\right)} \!: \ ,
\end{equation}
follows which clearly isolates the $z$-dependent part of the integrand. The
leading Regge term can \cite{Brower:2006ea} be isolated from this
expression by Taylor expanding with respect to $y$ in the exponential,
\begin{equation}
:\!e^{\ii \hat{p}_1 X(y)}\!: \, :\! e^{\ii \hat{p}_2 X(0)} \!: \sim \left(y\right)^{2 \al p_1 p_2} :\! e^{\ii \left(p_1 + p_2\right) X(0)  + \ii z y q^{\mu} \left(\partial X_{\mu}(0)\right) + \mathcal{O}\left(y^2\right)} \! : \ .
\end{equation}
Of course, a reasoning is needed to argue that the sub-leading terms in $y$ can be ignored. To obtain the full amplitude the complete correlation function must be calculated and then integrated. One can of course at least formally integrate the above expression directly. From a saddle-point approximation in the expression above a saddle point $y \sim \frac{1}{z}$ is obtained or in other words an effective expansion in terms of $(y z)$. For the purposes of this article this argument will be taken to hold - a more rigorous treatment would be very welcome as this will be our main assumption in this section.

Up to the order indicated the integral over the position of the first particle in the scattering amplitude from $-\infty$ to $0$ can be performed explicitly,
\begin{align}\label{eq:effecoper}
\int_{-\infty}^0 \left(y\right)^{2 \al p_1 p_2} & :\! e^{\ii \left(p_1 + p_2\right) X(0)  + \ii z y q^{\mu} \left(\partial X_{\mu}(0)\right)} \! : \, = \nonumber \\
& :\! e^{\ii \left(p_1 + p_2\right) X(0) } \left(\ii z q^{\mu} \left(\partial X_{\mu}(0)\right) \right)^{-1 -2 \al p_1 p_2} \Gamma(1+2 \al p_1 p_2)\! :  \ ,
\end{align}
with somewhat abstract reality conditions
\begin{equation}
\Re \left(\al p_1 p_2  \right) > -1 \quad \quad \Re (z q \left(\delta X(0)\right)) > 0  \ .
\end{equation}
Interestingly the effective operator in eq.~\eqref{eq:effecoper} again looks like a vertex operator for a physical state, apart from the non-familiar exponential. It is not too hard too verify that the operator indeed obeys the physical state conditions as
\begin{equation}
:\! e^{\ii \left(p_1 + p_2\right) X(0) } \left(\ii z q^{\mu} \left(\partial X_{\mu}(0)\right) \right)^{-1 -2 \al p_1 p_2}\! : |0 \rangle \sim \left(q_{\mu }\alpha_{-1}^{\mu}\right)^{-1 -2 \al p_1 p_2} | p_1 + p_2 \rangle  \ ,
\end{equation}
This is very similar to the vector excitation in the bosonic string. The physical state conditions reads $q \cdot (p_1 + p_2) =0$ which indeed holds. Note that this argument only holds for the leading term in the large $z$ expansion: sub-leading terms will spoil this behavior.

From the above expression the large $z$ behavior of the full amplitude can easily be isolated for the tachyon amplitude in the bosonic string,
\boxit{
\begin{equation}
A_n(z) \rightarrow \left(\frac{1}{z}\right)^{\al (p_1 + p_2)^2 - 1} \left( \tilde{G}_0 + \mathcal{O}\left(\frac{1}{z}\right) \right)  \ .
\end{equation}}
The CFT argument reproduces the adjacent shift behavior found in eq.~\eqref{eq:tachyonreccondition} for adjacent tachyons. Crucially, the evaluation of all remaining contractions in the vacuum expectation value can not depend on $z$. Note that in the CFT argument the shift behavior is manifestly universal: it only depends on the OPE of the shifted vertex operators.

Note also that since the leading term of the large $z$ expansion obeys the physical state condition, it obeys the same braiding relation as in eq.~\eqref{eq:braidingsignfactor}. This validates our argument above that non-adjacent shifts are suppressed by one additional power of $z$ compared to the leading one.

\subsection{Shifting gluons}
The argument above for tachyons can be extended to more general cases such as the shift of adjacent gluons. First this will be analyzed for the bosonic string, after which the RNS formulation of the superstring will be treated. In this subsection we will set $\al = \frac{1}{2}$, restoring dependence on this parameter by dimensional analysis at the end.

\subsubsection{Bosonic string}
In the bosonic string the vertex operators for gluons can be written as the part of
\be
\label{gluonzope}
:\! e^{\ii p_1\cdot X(y_1)+\gz_1\cdot \partial X(y_1)}\! :  \ ,
\ee
which is linear in the external polarization $\gz_1$. The exponential form is convenient for performing contractions, such as for instance for the OPE of two adjacent gluons,
\be
\label{gluonzope2}
:\! e^{\ii p_1\cdot X(y)+\gz_1\cdot \partial X(y)}\!: \, :\!e^{\ii p_2\cdot
  X(0)+\gz_2\cdot \partial X(0)}\!: \, \quad\sim\quad :\!K\,\,e^W \!: \ ,
\ee
where in the last line $K$ is the part of the OPE linear in both polarization vectors,
\be
\label{eq:K}
K=
  \gz_1\cdot \partial X(y) \gz_2 \cdot \partial X(0) +\frac{\gz_1\cdot
  \partial X(y) \gz_2 \cdot p_1 - \gz_1 \cdot p_2 \gz_2\cdot \partial
  X(0)}{y}  +\frac{\gz_1\cdot\gz_2 + \gz_1\cdot p_2 \gz_2\cdot p_1 }{y^2} \ ,
\ee
and $W$ is the same exponential as in the tachyon case (see eq.~\eqref{eq:afetrtachope})
\be
\label{eq:W}
W= p_1\cdot p_2 \log y +\ii \left(p_1\cdot X(y) +p_2\cdot X(0)\right) \ .
\ee
Note that up to this point everything is exact. Now the BCFW shift can be applied as above in eq.~\eqref{shift_rules}. The exponential will be Taylor expanded as
\begin{equation}
W \sim  p_1 p_2 \log y + \ii \left(p_1 + p_2\right) X(0)  + \ii z y q^{\mu} \left(\partial X_{\mu}(0)\right) + \mathcal{O}\left(y^2\right) \ ,
\end{equation}
To leading order the integral over $y$ can now be performed explicitly in the same way as above. Therefore for the shift of two adjacent gluons in any open bosonic string amplitude
\boxit{\begin{multline}\label{eq:masterexprglshCFT}
A_n(z) =  \left( \ov{z}\right)^{2\al p_1 p_2}  \hat{\gz}_1^\mu \left[ \int_{\textrm{positions}} \langle \Bigg\{z  \left[\eta_{\mu\nu} + 2\al P_{\mu}  P_{\nu} \right]  G_1 + \right. \\ \left.
   \left[\partial X(0) _\mu  P_{\nu} - P_{\mu} \partial X(0)_\nu\right] G_2   + \mathcal{O}\left(\ov{z} \right)  \Bigg\} V_3 V_4 \ldots V_n \rangle\right]\hat{\gz}_2^\nu  \ ,
\end{multline}}
holds, where explicit $\al$ has been restored. To obtain this expression
\begin{equation}\label{eq:osdorpposse}
p_1 \cdot \gz_2 = - P \cdot \gz_2 \qquad  \qquad p_2 \cdot \gz_1 = - P \cdot \gz_1  \ ,
\end{equation}
with
\begin{equation}
\sum_{i=3}^{n} p_i \equiv P  \ ,
\end{equation}
was used to eliminate any apparent $z$ dependence in the shifted momenta. The integration in eq.~\eqref{eq:masterexprglshCFT} is over the position of particles $3$ through $n-2$. The $G_i$ are operator dependent polynomials in $(\ov{z})$ with constant term. The form of eq.~\eqref{eq:masterexprglshCFT} is exactly the same as the leading and sub-leading terms in eq.~\eqref{eq:Vleadibosstr}. Antisymmetry of the sub-leading term is manifest. Hence the CFT argument reproduces the large shift behavior of Table \ref{tab:kinconbos}. Again, the CFT argument clearly indicates that the obtained behavior is independent of the particle content of the rest of the amplitude.

Again, the leading term of the large $z$ expansion obeys the physical state condition as above as the leading operator is simply the same. This verifies the claim that non-adjacent shifts are suppressed by one additional power of $z$ compared to the leading one as the full leading behavior is given by this term.

\subsubsection{Superstring in the RNS formulation}
The vertex operators of the superstring in the RNS formulation can be written as a double Grassmanian integral,
\begin{equation}
V(p, \gz, y) = \int d\theta d\eta V(\theta, \eta) = \int d\theta d\eta e^{\ii p_{\mu} X^{\mu}(y) + \eta \theta \gz_{\mu} \dbar X^{\mu}(y) + \theta p_{\mu} \psi^{\mu}(y) + \eta \gz_{\mu} \psi^{\mu}(y)}  \ .
\end{equation}
The calculation can now be set up completely analogously to the reasoning followed above. Let us first calculate the OPE,
\begin{multline}
:\!V_1(p_1, \gz_1, y_1)\!: \, : \!V_2(p_2, \gz_2, y_2)\!: =  \int d\theta_1 d\eta_1 d\theta_2 d\eta_2 :\!V_1(\theta_1, \eta_1) V_2(\theta_1, \eta_1)\!: \\
\exp\left[ p_1 p_2 \left(\log(y_2-y_1) + \frac{\theta_1 \theta_2}{y_2 - y_1}  \right)  + \eta_1 \theta_1  \eta_2 \theta_2 \frac{\gz_1 \gz_2}{(y_2-y_1)^2} - \eta_1 \theta_1 \frac{\gz_1 p_2}{y_2-y_1} + \right. \\ \left. \eta_2 \theta_2 \frac{p_1 \gz_2 }{y_2-y_1}  + \theta_1 \eta_2 \frac{p_1 \gz_2}{y_2 - y_1} + \eta_1 \theta_2  \frac{\gz_1 p_2}{y_2 - y_1} + \eta_1 \eta_2 \frac{\gz_1 \gz_2}{y_2 - y_1}  \right]  \ .
\end{multline}
The integrals over the Grassmanian parameters can be performed, which yields the same structure as before,
\begin{equation}
:\!V_1(p_1, \gz_1, y_1)\!: \, :\! V_2(p_2, \gz_2, y_2)\!: = :\!K_{\textrm{susy}} e^{W}\!:  \ ,
\end{equation}
with $W$ as in eq.~\eqref{eq:W}. The polarization dependent structure can be written as
\begin{equation}
K_{\textrm{susy}} = K_0 + \frac{K_1}{(y_2 -y_1)^1} + \frac{K_2}{(y_2 -y_1)^2}  \ ,
\end{equation}
with
\begin{align}
K_2 & = \left(\gz_1 \cdot \gz_2\right) (1 +  p_1 p_2) \\
K_1 & = (\gz_1 P) \left[ \gz_{2, \mu} \dbar X^{\mu}(y_2) + \gz_{2,\mu}  \psi^{\mu}(y_2) \left( p_{2,\nu}  \psi^{\nu}(y_2) + p_{1,\nu}  \psi^{\nu}(y_1) \right)\right] -  \nonumber \\
    & \quad \quad (\gz_2 P) \left[ \gz_{1, \mu} \dbar X^{\mu}(y_1) + \gz_{1,\mu}  \psi^{\mu}(y_1) \left( p_{1,\nu} \psi^{\nu}(y_1) + p_{2,\nu} \psi^{\nu}(y_2) \right)\right] +  \nonumber \\
    & \quad \quad \quad\quad(\gz_1 \gz_2) \left[p_{1,\mu} p_{2,\nu} \psi^{\mu} (y_1) \psi^{\nu}(y_2)\right] +  (p_1 p_2) \left[\gz_{1,\mu} \gz_{2,\nu} \psi^{\mu} (y_1) \psi^{\nu}(y_2)\right] \label{eq:KK1factor} \\
K_0 & =  \left[\gz_{1, \mu} \dbar X^{\mu}(y_1) + \gz_{1,\mu} p_{1,\nu}  \psi^{\mu}(y_1)\psi^{\nu}(y_1) \right] \left[\gz_{2, \mu} \dbar X^{\mu}(y_2) + \gz_{2,\mu} p_{2,\nu}  \psi^{\mu}(y_2) \psi^{\nu}(y_2) \right]  \ ,
\end{align}
where eq.~\eqref{eq:osdorpposse} was used. Compared to the bosonic string the analysis is more complicated since the above expressions contain explicit dependence on $z$ through $p_1$ and $p_2$. The leading order behavior of the amplitude follows from the above reasoning. The simplest is $K_2$. This contributes
\begin{equation}
A(z)_{\textrm{contribution from} K_2} \sim \left( \ov{z}\right)^{ p_1 p_2-1} \left[ \left(\gz_1 \cdot \gz_2\right) (1 +  p_1 p_2) \right]  \ .
\end{equation}
The other two are suppressed with respect to this contribution, $K_1$ by $\frac{1}{z}$ and $K_0$ by $\frac{1}{z^2}$. The strategy now will be to show that leading terms will be proportional to metric contractions between $\gz_1$ and $\gz_2$, while the sub-leading terms are antisymmetric in these two polarizations.

Expanding the integrand in $(y_2 - y_1)$ and repeating the integral argument shows that the leading behavior arises from the $K$ factors, evaluated at $y_2 = y_1$. Sub-leading contributions arise from higher order terms in the Taylor expansion in $(y_2-y_1)$. The first two lines of $K_1$ are explicitly anti-symmetric. Furthermore, the potentially dangerous $z$-dependence in this factor from the momenta cancels explicitly at leading order in $(y_2-y_1)$. The last line of $K_1$ contains one term proportional to the metric which due to the antisymmetry of $\psi \psi$ at leading order in $(y_2-y_1)$ can only contribute at the same order in $z$ as the contribution coming from $K_2$. The other term on the last line of $K_1$ is explicitly anti-symmetric at leading order in $(y_2 - y_1)$.

The remaining $K_0$ contribution can only contribute at leading or sub-leading order in $z$ if the $z$ dependence in the momentum would play a role. At leading order in $(y_2-y_1)$ the potentially dangerous terms come from
\begin{multline}
\sim \gz_{1,\mu} \hat{p}_{1,\nu}  \psi^{\mu} \psi^{\nu} \gz_{2,\kappa} \hat{p}_{2,\rho}  \psi^{\kappa} \psi^{\rho}  + \gz_{1, \mu} \dbar X^{\mu}  \gz_{2,\mu} \hat{p}_{2,\nu}  \psi^{\mu} \psi^{\nu} + \gz_{2, \mu} \dbar X^{\mu} \gz_{1,\mu} \hat{p}_{1,\nu}  \psi^{\mu}\psi^{\nu} = \\
z \gz_{1,\mu}\left[  \psi^{\kappa}  \psi^{\mu} \left(p_{1,\nu} q_{\rho} - q_{\nu} p_{2,\rho}   \right) \psi^{\nu} \psi^{\rho} +  \left(\dbar X^{\mu} (q_{\nu} \psi^{\nu})\psi_{\kappa} -  \dbar X^{\kappa} (q_{\nu} \psi^{\nu})\psi_{\mu}  \right) \right]\gz_{2,\kappa}  + \mathcal{O}\left(z^0\right)  \ ,
\end{multline}
where the shift has been implemented. These contributions are explicitly anti-symmetric while it can only contribute one order in $z$ higher: at the same level as $K_1$. At sub-leading order in $(y_2-y_1)$ the only possible non-zero contribution is anti-symmetric in $\gz_1$ and $\gz_2$ since the derivative has to act on one of the fields contracted into $q_{\mu} \psi^{\mu}$.

Gathering all the terms it is seen that the leading order behavior of the shift of two adjacent gluons in the superstring can be derived from
\boxit{\begin{multline}
A_n \sim \left(\frac{1}{z}\right)^{2 \al (p_1 \cdot p_n)}  \hat{\gz}_1^{\mu} \left( z \, h_1 (1+2 \al p_1 \cdot p_n) g_{\mu\nu} +  \left(h_2 B_{\mu\nu} \right) + \mathcal{O}\left(\frac{1}{z} \right) \right) \hat{\gz}_2^{\nu}  \ ,
\end{multline}}
where $B$ is an anti-symmetric tensor and $h_i$ are some polynomials in $\ov{z}$ with constant finite part. They can be related to certain operators in the CFT as before. In this expression explicit $\al$ has been restored. The derived formula allows one to reproduce Table \ref{tab:kinconsup} from the CFT argument. Just as above, the argument does not depend on the field content of the rest of the amplitude.

As above, for non-adjacent shifts our suppression argument will work as in the bosonic string: the braiding factor is the same. However, here the leading behavior in $z$ is in some cases determined from parts of the effective vertex operator which are either absent or sub-leading in the bosonic string. This happens in those cases which are better behaved compared to the bosonic string. The importance of the sub-leading terms makes it harder to prove the physical state condition in these cases for the superstring. We conjecture, but do not prove, that the same conclusion as there holds: suppression by one power of $z$. Note that this is certainly true in the field theory limit and can be checked to hold for the purely bosonic part.

\subsection{Remarks}
The CFT argument is seen to yield the same answers as the direct integral analysis as presented in section \ref{sec:proofopenstringrecadj}, at least in the cases checked here. The CFT is however remarkably more efficient. Also, in the CFT picture it is obvious that the large $z$ behavior should indeed be universal in string theory in the sense explained above as it only depends on the OPE between the two shifted particles, not on any other leg. The large $z$ behavior of any choice of two species of particle in the open string should also follow most easily from the above CFT argument. This can be seen from the general shape of the DDF vertex operators in the bosonic string \cite{DelGiudice:1971fp}, as these can be thought of as products of massless vertex operators. For the superstring it would be interesting to see if the OPE argument for the NSR formulation above can be made more straightforwardly in any formulation of the superstring with manifest target space supersymmetry.

Furthermore, non-adjacent shifts are expected to be suppressed with respect to the adjacent shift by application of \eqref{eq:monodromyexampleone}in general: the leading behavior in $z$ features in effect a universal effective particle made of the adjacent shift of $1$ and $2$. It would be nice to have a general proof of the physical state conditions. This argument yields one power of $z$ suppression however.

\section{Sample applications of the CFT argument}
\label{sec:curvback}

\subsection{Open strings in a constant B-field background}
As mentioned in the introduction, a longer term
goal of the research program of which this paper is a part of, is to
understand string theory in more general backgrounds than the flat
one which was discussed up to now. There is a small  but non-trivial
list of explicit backgrounds for which string theory can be analyzed
completely. An interesting example in this list is open string
theory in a non-zero but constant B-field background (see the
seminal paper by  Seiberg and Witten \cite{Seiberg:1999vs} and
references therein).

The results of the present article for disc-level string scattering
can easily be generalized to that setting. The crucial observation
is that a generic CFT correlation function of open string vertex
operators in the nontrivial B-field background can be calculated
simply from the flat background case,
\begin{equation}\label{eq:commvsnoncomm}
\langle V_1 V_2 \ldots V_n \rangle_{B \neq 0}  =  e^{-\frac{\ii}{2} \sum_{i<j} p^{\mu}_i
p_j^{\nu} \theta_{\mu \nu} \epsilon(z_i, z_j)} \langle V_1 V_2
\ldots V_n \rangle_{B =0, \eta_{\mu\nu} \rightarrow G_{\mu\nu}} \ .
\end{equation}
In this formula the non-commutativity parameter $\theta$ is defined
as
\begin{equation}
\theta^{\mu \nu}  =  - (2\pi \al)^2 \left(\frac{1}{g + 2 \pi \al B } B \frac{1}{g - 2 \pi \al B
}\right)^{\mu \nu} \ ,
\end{equation}
where $g$ is the (flat) closed string metric and $B$ the background
B-field. Furthermore, the flat space metric $\eta_{\mu\nu}$ is replaced by the
open string metric,
\begin{equation}\label{eq:defopenstrmetr}
G^{\mu \nu}  = \left(\frac{1}{g + 2 \pi \al B } g \frac{1}{g - 2\pi \al B
}\right)^{\mu \nu} \ .
\end{equation}
In the field theory limit described in \cite{Seiberg:1999vs} the
target space theory reduces to the non-commutative version of
Yang-Mills theory. In the limit where $B\rightarrow 0$, the flat
space correlation functions are recovered. The string amplitudes
follow from the correlation function in the usual way by integration
over insertion points as in \eqref{eq:genopenstringamp}.

\subsubsection*{Monodromy relations}
From the relation to the regular string theory amplitudes it is
clear that monodromy relations can be derived for the amplitudes in
a non-trivial B-field background. More elegantly, this can be done
by deriving the co-cycle factor from the string theory OPE and
following the derivation of the monodromy relations above. This
yields
\begin{equation}\label{eq:braidingsignfactornoncom}
:\!V_1(z_1)\!:\, :\!V_2(z_2)\!: \equiv :\!V_2(z_2)\!:\, :\!V_1(z_1)\!: e^{ \ii
p^{\mu}_1 p^{\nu}_2 \left(2 \pi \al G_{\mu \nu} - \theta_{\mu\nu}\right)
\epsilon({z_1,z_2})} \ .
\end{equation}
Since the co-cycle factor does not depend on the insertion points
the derivation in subsection \ref{sec:monfromcft} can be copied directly. This yields a first
monodromy relation between different color-orders as
\begin{equation}
A(\alpha_1,1,\beta_1 \ldots \beta_k , n)_B = -  \sum_{\sigma \in
OP\{\alpha_1\} \cup \{\beta\}} e^{\pm (\alpha_1 1)} \left(
\prod_{i=1}^k e^{\pm(\alpha_1 \beta_i)_{\theta}} \right)  A(1, \sigma, n)_B  \ .
\end{equation}
Here we have defined the inner product
\begin{equation}
( \alpha \beta)_\theta \equiv \ii \alpha^\mu \beta^\nu (2 \pi \al G_{\mu \nu} -
\theta_{\mu \nu}) \ ,
\end{equation}
which enters in the  definition  \eqref{eq:phasefac}. Note that similar constraints as above apply for reality conditions of the kinematic invariants for these relations to be valid. Iterating these relations yields the generalization of eq.~\eqref{eq:towardbjbdamvhresult},
\begin{multline}\label{eq:towardbjbdamvhresultnoncom}
A(\alpha_1 \ldots \alpha_r 1 \beta_1 \ldots \beta_s n) = \\(-1)^r
\left( \prod_{i} e^{\pm (\alpha_i 1)_\theta } \right) \left( \prod_{i>j}
e^{\pm (\alpha_i \alpha_j)_\theta} \right) \left( \sum_{\sigma \in
OP\{\alpha\}^T \cup \{\beta\}} \left( \prod_{i=1}^r \prod_{j=1}^{s}
e^{\pm (\alpha_i \beta_j)_\theta}\right) A(1 \sigma n)  \right) \ ,
\end{multline}

Using both possible choices of sign the amplitude can be ordered
further. This expresses any color ordered amplitude in terms of an
$(n-3)!$ element basis as in the flat space case. As for the flat space case, the monodromy relations hold for any possible field
content of the scattering amplitudes.

\subsubsection*{On-shell recursion in the presence of a B-field}
Recursion relations can be obtained for the string theory amplitudes using the techniques of either section \ref{sec:proofopenstringrecadj} or \ref{sec:CFTderivation}, modulo one subtlety with the phase factor in
\eqref{eq:commvsnoncomm}. For a generic BCFW shift this phase factor
will diverge wildly. However, the divergence is well-localized in
any color ordered amplitude. Shifting particles $i$ and $i+1$, this
takes the form
\begin{equation}
\sim e^{z q^{\mu} \theta_{\mu\nu} \left(p_i + p_{i+1} \right)^{\nu} } \
.
\end{equation}
In other words, consider
\begin{equation}
\label{Brec}
\oint_{z=0} e^{- z q^{\mu} \theta_{\mu\nu} \left(p_i + p_{i+1}
\right)^{\nu} } \frac{A(z)}{z} \ ,
\end{equation}
instead of eq.~\eqref{eq:BCFWstarting} as the starting point of the derivation of the BCFW relations.  As the exponential is an entire function, this does not introduce additional poles. The behavior of the string amplitudes in the large-$z$ limit is given by the flat space results of Tables \ref{tab:kinconbos} and \ref{tab:kinconsup}, replacing the target space metric by the open string metric $G$ in eq.~\eqref{eq:defopenstrmetr}. The analysis is completely parallel, up to and including the fact that only one monodromy relation is valid for a given non-adjacent large-$z$ shift. For this shift the amplitude decays exponentially. Note that the resulting recursion relation simply instructs one to calculate the $B=0$ amplitudes through BCFW recursion, replace the metric by the open string metric and multiply by the phase factor. This reproduces the prescription described in \cite{Raju:2009yx}.

\subsection{Closed strings at the level of the sphere}
For string theory in a flat background the vertex operators for the closed string sector are products of left and right moving vertex operators,
\begin{equation}
V^{\textrm{cl}}\left(z, \bar{z},\{p, \xi\} \right)= V_L^{\textrm{o}}\left(z, \{ p, \gz^L\}\right) V_R^{\textrm{o}} \left(\bar{z} \{ p, \gz^R\}\right)  \ .
\end{equation}
Here on the left hand side the $\xi$ are the polarization tensors of the closed string modes, while on the right hand side within the color ordered amplitudes the $\zeta^{L}$ and $\zeta^{R}$ are the polarization tensors of the left and right moving open string modes respectively. These two are related by
\begin{equation}
\xi = \zeta^L \otimes \zeta^R  \ ,
\end{equation}
where the symbol $\otimes$ may include algebraic restrictions. For instance for gravitons this simply expresses the graviton polarization tensor in terms of the traceless symmetric product of vector polarizations,
\begin{equation}
h_{\mu\nu} = \frac{1}{2}\left(\gz^R_{\mu} \gz^L_{\nu} + \gz^R_{\nu} \gz^L_{\mu} \right) - \frac{\eta^{\mu\nu}}{D} \left(\gz^R \cdot \gz^L\right)  \ .
\end{equation}
The above simply shows that for free field theory the spin $2$ representation can be found in the tensor product of $2$ spin $1$ representations. This squaring relation seems special to free field theory. However, as shown by KLT \cite{Kawai:1985xq} a general relation between closed and open string amplitudes was uncovered which generalizes the free field 'squaring' relation.

As an example first study four, five and six point amplitudes. The main assumption will be that the KLT relations hold in the form derived in \cite{Kawai:1985xq} also in the large $z$ limit. Regardless of particle content the KLT relations can be taken to read for four particles,
\begin{multline}
A_4^{\textrm{cl}}\left(\{p_i, \xi_i\} \right) = \sin\left(2 \pi \al t\right) \,\, A_L^{\textrm{o}}\left(\{p_1, \gz^L_1\}, \{p_2, \gz^L_2\},\{p_3, \gz^L_3\},\{p_4, \gz^L_4\} \right) \\
A_R^{\textrm{o}}\left(\{ p_1, \gz^R_1\},\{p_3, \gz^R_3\},\{p_2, \gz^R_2\},\{p_4, \gz^R_4\} \right)  \ .
\end{multline}
Hence shifting particles $1$ and $4$ on the closed string amplitude gives immediately the product of the shifts of the two open string amplitudes. If the shifts of the open string amplitudes are known, a precise prediction for the large $z$ behavior is obtained. Note that a shift of particles $1$ and $2$ would lead to the same prediction after using the non-adjacent shift analysis above. For five particles KLT give
\begin{multline}
A_5^{\textrm{cl}}\left(\{p_i, \xi_i\} \right) \sim  \sin\left(2 \pi \al p_1 \cdot p_2 \right) \sin\left(2 \pi \al p_3 \cdot p_4 \right) \,\, A_L^{\textrm{o}}\left(13245\right) A_R^{\textrm{o}}\left(31425\right) + \\ \sin\left(2 \pi \al p_1 \cdot p_3 \right) \sin\left(2 \pi \al p_2 \cdot p_4 \right) \,\, A_L^{\textrm{o}}\left(12345\right) A_R^{\textrm{o}}\left(21435\right)  \ ,
\end{multline}
where the tensor structure in the open string amplitudes has been suppressed for notational clarity on the right hand side. Shifting particles $1$ and $5$ shows as above a nice cancelation of the exponential factors between the sine functions and the amplitudes through the use of the non-adjacent shift of eq.~\eqref{eq:gennonadjshift}. A similar conclusion holds for the six-point formula given in \cite{Kawai:1985xq}. For arbitrary multiplicity KLT give
\begin{equation}
A_n^{\textrm{cl}}\left(\{p_i, \xi_i\} \right) \sim \sum_{P,P'} A_L^{\textrm{o}}\left(P\right) A_R^{\textrm{o}}\left(P\right) e^{\ii \pi F(P,P')}  \ ,
\end{equation}
where the sum is over all permutations $P$ and $P'$ of $1, \ldots n-1$ such that $1$ appears before $n-1$. The phase factor is  determined from the ordering between pairs of particles in $P$ and $P'$
\begin{equation}
F(P,P') = \sum_{i,i'=1}^{n-1} f(i,i')  \ ,
\end{equation}
with
\begin{equation}
f(i,i') = \left(\begin{array}{ccl} 0  & &\textrm{if particles } \{i,i'\} \textrm{ same ordering in } \{P,P'\} \\
                                2 \pi \al p_i \cdot p_{i'} & &\textrm{if particles } \{i,i'\} \textrm{ different ordering in } \{P,P'\} \end{array} \right.  \ .
\end{equation}
Again assuming that the derivation of the KLT relations is unaltered by the shift, the same conclusion as above follows: the large $z$ behavior of the closed string amplitude is the square of the open string one. Indeed, the only thing which could spoil this is the phase factor which can diverge under a large $z$ shift. Without loss of generality, let us shift particles $1$ and $n$. The only source for a potential divergence is then the case where particle $1$ appears with opposite order in the left and right sector open string amplitudes compared to some other particle, say $j$. Schematically this contribution reads
\begin{equation}
\sim e^{\ii 2 \pi \al p_i \cdot p_1} A(\ldots 1 \ldots j \ldots n) A(\ldots j \ldots 1 \ldots n)  \ .
\end{equation}
The divergent phase is then canceled by the phase factor arising from the non-adjacent shift derived above in eq.~\eqref{eq:gennonadjshift}. This reasoning can be repeated for any particle, completing the derivation. Let us stress that this argument relies on the validity of the KLT relations under the shift. Even disregarding applications to recursions, it would be interesting to find a completely CFT based derivation of the KLT relations along the lines of the derivation of the monodromy relations discussed above.


\subsubsection*{Shifting gravitons}
Up to now these results hold for any Lorentz quantum numbers of the external states. It is interesting to specialize to the case of gravitons. Combining our results for the gluonic amplitudes (tables \ref{tab:kinconbos} and \ref{tab:kinconsup}) tables \ref{tab:kinconbosclosed} and \ref{tab:kinconsupclosed} are obtained for the shift of two gravitons in the bosonic and superstring cases. These tables are again claimed to be universal and in particular to hold for all multiplicities. Again, the different possibilities for the transverse polarization are related to whether or not the different transverse polarizations in the problem are orthogonal.

\begin{table}[!h]
\begin{center}
\begin{tabular}{c|c c c c c c}
$\xi_1 \;\backslash \;\xi_n  $ &  $--$ &  $-+$ &  $++$ &  $-$T &  $+$T &  TT   \\
\hline
$--$                           & $ -2$ & $  0$ & $ +2$ & $  0$ & $ +2$ & $ +2$ \\
$-+$                           & $ -4$ & $ -2$ & $  0$ & $ -2$ & $  0$ & $  0$ \\
$++$                           & $ -6$ & $ -4$ & $ -2$ & $ -4$ & $ -2$ & $ -2$ \\
$-$T                           & $ -2$ & $  0$ & $ +2$ & $ -2/-1$ & $ 0/+1$ & $ 0/+1$ \\
$+$T                           & $ -4$ & $ -2$ & $  0$ & $ -3/-4$ & $ -2/-1$ & $ -2/-1$ \\
TT                             & $ -2$ & $  0$ & $ +2$ & $ -2/-1$ & $ 0/+1$ & $ -2/-1/0$
\end{tabular}
\caption{Conjectured leading power in $z^{- 2 \al(p_1+p_n)^2 -\kappa}$ for the large $z$ limit of the shift of two gravitons in the \textbf{bosonic string} for all possible polarizations.
  \label{tab:kinconbosclosed}}
\end{center}
\end{table}

\begin{table}[!h]
\begin{center}
\begin{tabular}{c|c c c c c c}
$\xi_1 \;\backslash \;\xi_n  $ &  $--$ &  $-+$ &  $++$ &  $-$T &  $+$T &  TT   \\
\hline
$--$                           & $ +2$ & $ +2$ & $ +2$ & $ +2$ & $ +2$ & $ +2$ \\
$-+$                           & $ -2$ & $ -2$ & $ +2$ & $ -2$ & $  0$ & $  0$ \\
$++$                           & $ -6$ & $ -2$ & $ +2$ & $ -4$ & $  0$ & $ -2$ \\
$-$T                           & $  0$ & $  0$ & $ +2$ & $0/+1$ & $ 0/+1$ & $ 0/+1$ \\
$+$T                           & $ -4$ & $ -2$ & $  0$ & $ +1$ & $  0/+1$ & $ -2/-1$ \\
TT                             & $ -2$ & $  0$ & $ +2$ & $ +1$ & $  0/+1$ & $ -2/-1/0$
\end{tabular}
\caption{Conjectured leading power in $z^{- 2 \al (p_1+p_n)^2 -\kappa}$ for the large $z$ limit of the shift of two gravitons in the \textbf{superstring} for all possible polarizations.
  \label{tab:kinconsupclosed}}
\end{center}
\end{table}

As an argument for tables \ref{tab:kinconbosclosed} and \ref{tab:kinconsupclosed}, consider the closed string version of the CFT argument for the large z-behavior explored above. This follows from \cite{Brower:2006ea} quite directly. For the OPE one considers
\begin{equation}
:\! V^{\textrm{cl}}\left(w_1, \bar{w}_1 \right)\!: \,:\! V^{\textrm{cl}}\left(w_2, \bar{w}_2 \right)\!:= \left[:\! V_L^{\textrm{o}}\left(w_1\right)  \!: \, :\! V_L^{\textrm{o}}\left(w_2\right)  \!: \right] \, \left[:\! V_R^{\textrm{o}} \left(\bar{w}_1\right) \!:\,:\! V_R^{\textrm{o}} \left(\bar{w}_2\right) \!: \right]  \ .
\end{equation}
The saddle point in the integral is at $z (w_1-w_2) \sim z (\bar{w}_1-\bar{w}_2) \sim 1$. Fixing $w_2 = 0 = \bar{w}_2$, the remaining integral over $w_1$ and $\bar{w}_1$ over the full complex plane can be performed on the formal level at least by analytic continuation to disentangle both integrals. However, this will yield simply the product of the open string analysis in both left and right sector performed above. As in the open string case, a full justification of the CFT argument will not be given here.

In the case of shifted gravitons the CFT argument leads to the tables \ref{tab:kinconbosclosed} and \ref{tab:kinconsupclosed} in the bosonic and superstring respectively. The worse behavior of the bosonic string amplitudes compared to the superstring is just as in the gluon case down to terms proportional to $\al$. Taking this into account, both have a smooth field theory limit. In the field theory the terms proportional to $\al$ which appear in the bosonic string probably correspond to the known $\al R^2$ and $\al^2 R^3$ terms in the effective action which are known to be absent in the superstring. A direct field theory argument here seems difficult however.

\section{Discussion and conclusion \label{sec:conc}}
In a precise sense, the recursion relations explored in this article show that the integrand in the string theory amplitude is a complete derivative: the full string amplitude can be determined from the singular behavior on the boundary of the moduli space of the amplitude. Interestingly, for a chosen shift only part of the boundary behavior is necessary. Different BCFW shifts which use different parts of the boundary lead to different expressions for the same amplitude. These different expressions are all equivalent, a fact which has recently been discussed in \cite{Hodges:2009hk} and \cite{ArkaniHamed:2009dn} from the field theory point of view. In string theory, this is not a bug but a feature. The equivalence of these different expressions is an exact expression of old-school Dolen-Horn-Schmid \cite{Dolen:1967jr} `duality' (sometimes also called crossing symmetry). The proof of the recursion relations depends on the complexified high energy Regge behavior of a pair of particles. On-shell recursion therefore seems to be a consequence of the `high energy symmetry algebra' of the string theory \cite{Gross:1988ue}.

Knowledge of the full string three-point amplitude and the on-shell recursion relation suffice to calculate higher point amplitudes in a flat background at least in principle. In a sense, the recursion relations must be an associativity condition in a topological theory while the three-point functions are structure constants, perhaps along the lines of \cite{Brooks:1992pg}. Further thought along these lines might lead one to investigate string field theory, especially Witten's cubic open bosonic string field theory. Note that off-shell Berends-Giele \cite{Berends:1987me} type recursion relations (which involve a current instead of an amplitude) have been studied for open string field theory in \cite{Sakai:1990ym} and for closed string field theory in \cite{Ilderton:2006hm}. Relating supersymmetric string field theory to the on-shell recursion relations should be very interesting since on-shell recursion avoids any problems with contact terms.

As a tool for practical calculation the on-shell recursion relations are not effective as of yet. Although in principle three particle interactions are fixed by the usual string analysis, especially at higher mass levels the resulting three-point functions are in their current form not very transparent for our purposes (See e.g. \cite{DiVecchia:1986uu} and references therein). It would be very interesting to study these using the four dimensional (massive) spinor helicity formalism or even better its higher dimensional extension \cite{Boels:2009bv} (see also \cite{Cheung:2009dc}). It would also be interesting to use the higher dimensional supersymmetric coherent states constructed in \cite{Boels:2009bv} to apply the higher dimensional version of the super-shift \cite{ArkaniHamed:2008gz} to string amplitudes. In four dimensions this can be done straightforwardly for massless particles by application of the results of this article. It would also be very interesting to study the interplay between the large $z$ behavior and the $\al$ expansion.

There are several broader directions in which the research presented in this article could be pursued further.

\subsubsection*{General backgrounds?}
Since the above is (the beginning of) a CFT understanding of on-shell recursion and a real derivation of amplitude relations solely from the CFT, an immediate question is whether or not these observations generalize to string theories in a different setting such as in other backgrounds. One example of this is a constant Abelian vector field background, which will feature in \cite{Boels:2010RH}. More generally, it would be highly interesting to study the AdS background in the superstring: the OPE tool developed above was after all inspired by \cite{Brower:2006ea} which mainly focused on Regge behavior in the curved background.

In this vein we would like to make an intriguing observation. In general the open string vertex operators obey a braiding relation,
\begin{equation}
:\!V_1(z_1)\!:\,\, :\!V_2(z_2)\!: \,\,\equiv \,\, :\!V_2(z_2)\!:\, :\!V_1(z_1)\!:\, R_{12}  \ ,
\end{equation}
which is the generalization of the flat background relation \eqref{eq:braidingsignfactor} where the matrix $R_{12}$ is simply a phase. Now it is known since \cite{Tsuchiya:1988fy,Moore:1989ni} that consistency of the three point function requires the matrix $R_{12}$ to obey the Yang-Baxter equation. This is trivially true in a flat background, see e.g. \cite{Halpern:1996et} for examples in non-flat backgrounds. That relation was crucial in the derivation of the amplitude relations for different color orders, but apart from this the only ingredient was the generic structure of the integration domains. This observation provides a direct link of amplitudes to integrability issues, and it would be very interesting to explore this link further.

\subsubsection*{String loops?}
The CFT understanding of recursion also begs for the question if similar observations can be extended to the string loop level. In fact, from the CFT argument it is at least `hand-wavingly obvious' this is the case, and it would be exceedingly interesting to make this more precise. The main difference to the tree level case is the appearance of branch cuts in the $z$ plane, in addition to closed string poles. An analyticity argument could be used to excise certain points from the contour at infinity at the tree level, but at loop level this is no longer true. Instead, one obtains integrals around the branch cuts which measure the discontinuity across the branch cut. This can be related to amplitudes with a lower amount of loops or legs at least in principle. Note the close relation of this with the `sewing construction' for the correlation functions of any CFT at higher loops.

Loop level recursion in string theory would certainly be interesting with a view of applications to field theory, where it seems prohibitively difficult to obtain the large $z$ behavior in full generality for the (dimensionally regulated) amplitude. Based on a known field theory example in \cite{Bern:2005hs}, a tension can be expected between the various limits ($\al \rightarrow 0$, $D \rightarrow 4$ and $z \rightarrow \infty$ ) at least for the non-supersymmetric case.

\subsubsection*{An amplitude bootstrap?}
Related to the previous two points, the equivalence of the different recursion relations for string theory amplitudes at tree level is highly reminiscent of the bootstrap equations in conformal field theory. The analogy of possible recursion relations for amplitudes at loop level to the calculation of CFT correlation functions through the sewing construction was also mentioned above. The difference is that in CFT the bootstrap idea applies to the (un-integrated) correlation functions, while for the amplitudes these functions must be integrated over the moduli space. Still, it would be very interesting to make the analogy to the CFT bootstrap more precise. In its most extreme form it would give an equation for the S-matrix, the solution of which forms the space of all string theory amplitudes.

The B-field example however shows already that some care might be required in order to properly obtain the right formulae. In particular the modification of the derivation in eq.~\eqref{Brec} stands out here. A preliminary version of the bootstrap idea as applied to field theory was presented in \cite{Benincasa:2007xk}.

\subsubsection*{Finally,}
it is an encouraging sign that the recent spectacular progress in field theory can be matched by the string theory, at least as far as on-shell recursion for tree amplitudes is concerned. After all string theory should be an upgrade of field theory, not a downgrade. This article should be taken as inspiration to further study the fruitful interplay between string and field theory techniques.

\acknowledgments It is a pleasure to thank Jan de Boer and Emil Bjerrum-Bohr for discussions and comments. Our sole figure was made with JaxoDraw \cite{Binosi:2008ig}. The research of RHB is supported by a Marie Curie European Reintegration Grant within the 7th European Community Framework Programme.

\appendix

\section{Four-gluon amplitude in bosonic and superstring theory}
\label{fourgluons}

As an explicit example and cross-check on our results in this appendix BCFW shifts of the four-gluon amplitude
in bosonic and type I superstring theory will be discussed. Particles $1$ and $2$ will be shifted.

\subsubsection*{The four-gluon amplitude}

The four-gluon amplitude in the open bosonic strings can be found in for instance \cite{Kawai:1985xq}. Introducing standard Mandelstam
variables $s=-(p_1 + p_2)^2$, $t=-(p_1+p_4)^2$ and $u=-(p_1+p_3)^2$ and restoring $\al$ factors the amplitude can be written as
\begin{equation}
\label{four_gluons}
A_4(p_1,\gz_1;p_2,\gz_2;p_3,\gz_3;p_4,\gz_4)=\frac{\Gamma(-\al s
-1)\Gamma(-\al t -1)}{\Gamma(\al u +2)}
\cK(p_1,\gz_1;p_2,\gz_2;p_3,\gz_3;p_4,\gz_4) \ .
\end{equation}
Here $\cK$ is a cumbersome kinematic factor
\begin{multline}
\cK(p_1,\gz_1;p_2,\gz_2;p_3,\gz_3;p_4,\gz_4) =\\ (\al s+1)(\al
t+1)(\al u+1)\left[-K^{\text{(sup)}}+ K^{(s)}+
K^{(t)}+K^{(u)}+K^{(stu)} \right] \ ,
\end{multline}
{\footnotesize
\be
\begin{split}
K^{\text{(sup)}} &= -\al^2 (st\gz_1\cdot\gz_3 \gz_2\cdot\gz_4+su\gz_2\cdot\gz_3 \gz_1\cdot\gz_4+tu\gz_1\cdot\gz_2 \gz_3\cdot\gz_4)\\
&+2\al^2 s(\gz_1\cdot p_4 \gz_3\cdot p_2 \gz_2\cdot\gz_4 +\gz_2\cdot p_3 \gz_4\cdot p_1 \gz_1\cdot\gz_3+\gz_1\cdot p_3 \gz_4\cdot p_2 \gz_2\cdot\gz_3 + \gz_2\cdot p_4 \gz_3\cdot p_1 \gz_1\cdot\gz_4)\\
&+2\al^2 t(\gz_2\cdot p_1 \gz_4\cdot p_3 \gz_3\cdot\gz_1 +\gz_3\cdot p_4 \gz_1\cdot p_2 \gz_2\cdot\gz_4+\gz_2\cdot p_4 \gz_1\cdot p_3 \gz_3\cdot\gz_4 + \gz_3\cdot p_1 \gz_4\cdot p_2 \gz_2\cdot\gz_1)\\
&+2\al u^2(\gz_1\cdot p_2 \gz_4\cdot p_3 \gz_3\cdot\gz_2 +\gz_3\cdot
p_4 \gz_2\cdot p_1 \gz_1\cdot\gz_4+\gz_1\cdot p_4 \gz_2\cdot p_3
\gz_3\cdot\gz_4 + \gz_3\cdot p_2 \gz_4\cdot p_1 \gz_1\cdot\gz_2) \ ,
\end{split}
\ee
\be
\begin{split}
K^{(s)}&=4\al^3 s \Bigg\{ \gz_1\cdot p_3 \gz_2\cdot p_3(\gz_3\cdot p_1 \gz_4\cdot p_1 + \gz_3\cdot p_2 \gz_4\cdot p_2)+\ov{3}(\gz_1\cdot p_2\gz_2\cdot p_3 \gz_3\cdot p_1-\gz_1\cdot p_3 \gz_2\cdot p_1 \gz_3\cdot p_2) \\
&\times (\gz_4\cdot p_1 -\gz_4\cdot p_2)\Bigg\} \ ,
\end{split}
\ee
\be
\begin{split}
K^{(t)}&=4\al^3 t \Bigg\{\gz_2\cdot p_1 \gz_3\cdot p_1(\gz_1\cdot p_3 \gz_4\cdot p_3 + \gz_1\cdot p_2 \gz_4\cdot p_2)+\ov{3}(\gz_1\cdot p_3\gz_2\cdot p_1\gz_3\cdot p_2-\gz_1\cdot p_2 \gz_2\cdot p_3 \gz_3\cdot p_1) \\
&\times (\gz_4\cdot p_3 -\gz_4\cdot p_2) \Bigg\}\ ,
\end{split}
\ee
\be
\begin{split}
K^{(u)}&=4\al^3 u \Bigg\{\gz_1\cdot p_2 \gz_3\cdot p_2(\gz_2\cdot p_1 \gz_4\cdot p_1 + \gz_2\cdot p_3 \gz_4\cdot p_3)+\ov{3}(\gz_1\cdot p_2\gz_2\cdot p_3\gz_3\cdot p_1-\gz_1\cdot p_3 \gz_2\cdot p_1 \gz_3\cdot p_2) \\
&\times (\gz_4\cdot p_3 -\gz_4\cdot p_1) \Bigg\} \ ,
\end{split}
\ee
\be
\begin{split}
K^{(stu)}&=\al^2 st\left[\ov{\al u+1}(\gz_1\cdot\gz_3-2\al \gz_1\cdot
  p_3\gz_3\cdot p_1)(\gz_2\cdot\gz_4-2\al\gz_2\cdot p_4 \gz_4\cdot p_2)-(\gz_1\cdot\gz_3 \gz_2\cdot\gz_4)\right] \\
&+ ( 3\leftrightarrow 2) + (3\leftrightarrow 4 )\ ,
\end{split}
\ee
}

\noindent
The factor $K^{\text{(sup)}}$ is the same as it appears in the analogous type I superstring computation. Note that both $\cK$ and $K^{\text{(sup)}}$ are $stu$ symmetric so we do not need to compute  every term explicitly.

The contribution of the $\Gamma$ factor in (\ref{four_gluons}) coming from the shift $p_1\to p_1+zq\,,p_2\to
p_2-zq$ is easily derived by considering that (\ref{four_gluons}) itself is nothing but a shifted Beta function, as we already computed  in
(\ref{eq:openstringregge4pt}). There are however extra powers of $\al$
and $z$ here to take care of. They are crucial indeed as they give an
effective $\al z$ expansion
{\footnotesize
\begin{multline}
\label{eq:alexpansionbeta}
\frac{\Gamma(-\al s -1)\Gamma(-\al t -1)}{\Gamma(\al u +2)}=
\frac{\Gamma(-\al s -1)\Gamma(-\al t -1)}{(\al u -2)(\al u -1)(\al
u)(\al u +1)\Gamma(\al u -2)}\\
\sim \left(\ov{2 \al z q\cdot
  p_4} \right)^{\al s -1} \ov{(q\cdot p_4)^4\al^4 z^4 }
\left[1+\half \al z + \cO(\al^2 z^2)  \right]
\Gamma(\al s -1)  \ .
\end{multline}
}

\noindent
 Now, shifting particles 1 and 2 according to
(\ref{shift_rules}) we get the explicit
scaling behavior of the complete kinematic factor for any different
choice of the polarization vectors associated with shifted
momenta. To actually get the correct contributions, an expansion in
powers of $\al$ is needed as well. Sub-leading  contributions (in
$\al$), corresponding in the field theory limit to $\al F^3$ and
higher corrections, may also differ from the picture in Table
\ref{tab:kinconbos}. The ``good'' and ``bad'' shifts for the bosonic
string read

{\footnotesize
 \be
\cK^{+-} \sim -16\al^4 \left( 2\al^2(q\cdot q^*)^2 -\frac{\al s}{2}
\right)(q\cdot p_4)^4 \left( \gz_3\cdot\gz_4 - 2 \al  k_3\cdot\gz_4
  k_4\cdot\gz_3 \right) z^6+ \cO(z^5)  \ ,
\ee

\be
\cK^{-+} \sim 8 \al^4 \al s (\al s -1)(q\cdot p_4)^4 \left( \gz_3\cdot\gz_4 - 2 \al  k_3\cdot\gz_4
  k_4\cdot\gz_3 \right) z^2+ \cO(z)  \ .
\ee
}

\noindent
For $++$ and $--$ additional terms appear carrying a higher power of $z$
than expected. However note that these terms also carry additional
powers of $\al$. Anyway it does not seem to be an order of limit
problem, since the field theory limiting behavior of the
string theory amplitude is obtained by first taking the zero slope
limit and then sending $z\to\infty$. Higher corrections carry
extra $\al$ factors that do not cancel with the denominator of
(\ref{eq:alexpansionbeta}), this in turn guarantees that the
relations in Table
\ref{tab:kinconbos} are not spoiled.

{\footnotesize
\be
\begin{split}
\cK^{++}\sim &-4\al^4 s  (q\cdot p_4)^2 \left( q\cdot q^*
  \gz_3\cdot\gz_4 -2 p_2 \cdot\gz_3 p_2 \cdot\gz_4 \right) z^2\\
&+16s \al^6 z^4 (q\cdot p_4)^4 ( \gz_3\cdot\gz_4)+ \cO(\al^6 z^3)
\end{split}  \ ,
\ee

\be
\begin{split}
\cK^{--}\sim &-4\al^4 s  (q\cdot p_4)^2 \left( q\cdot q^*
  \gz_3\cdot\gz_4 -2 p_1 \cdot\gz_3 p_1 \cdot\gz_4 \right) z^2\\
&+16s \al^6 z^4 (q\cdot p_4)^4 ( \gz_3\cdot\gz_4)+ \cO(\al^6 z^3)
\end{split}  \ .
\ee
}

\noindent
For transverse polarization vectors we can pick both $\gz_1$ and
$\gz_2$ to be along the same transverse direction, indicated with $T$,
or chose different directions $T_1$ and $T_2$ so that
$\gz_1^{T1}\cdot\gz_2^{T2}=0$.
As in field theory, the latter shift is better behaved than the former
by one power of $z$.

{\footnotesize
\be
\cK^{TT}\sim 16 \al^4 z^4  (q\cdot p_4)^4 \gz_1^T\cdot\gz_2^T
(\gz_3\cdot\gz_4-2\al p_3\cdot\gz_4 p_4\cdot\gz_3) + \cO(z^3)  \ ,
\ee

\be
\cK^{T1T2}\sim 16 \al^4 z^3  (q\cdot p_4)^3
(p_3\cdot \gz_1^{T1}  p_4\cdot\gz_2^{2T}-p_4\cdot \gz_1^{T1}  p_3\cdot\gz_2^{2T})  \ .
\ee
}

\subsubsection*{Type I superstrings}

In the type I superstring case the four-gluon
 amplitude has the
same structure as (\ref{four_gluons}) with the restriction on the
kinematic coefficient being
\cite{Schwarz:1982jn,Kawai:1985xq}:

{\footnotesize
 \be \cK =
-(\al s+1)(\al t+1)(\al u+1)K^{\text{sup}} \ .
\ee
}

\noindent
The large $z$ scaling behavior of this amplitude can be computed the
same way as before and leads to

{\footnotesize

\be
\cK^{+-} \sim 8\al^5 s (q\cdot p_4)^4 (\al s +1)\gz_3\cdot\gz_4 z^6 + \cO(z^5)  \ ,
\ee

\be
\cK^{-+} \sim 8\al^5 s (q\cdot p_4)^4 (\al s +1)\gz_3\cdot\gz_4  z^2+ \cO(z)  \ .
\ee
}

\noindent
In both the $++$ and $--$ case there are some cancelations to take
into account to get the correct large $z$ behavior of the kinematic
factors. Such cancelations are due to our particular choice of  a
Lorentz frame where $s=2p_1\cdot p_2 =2$ and $q\cdot q^*=1$, however
this only implies a smart choice of unit of measure, which indeed is
always possible. Gathering factors that depend on $s-2q\cdot q^*$ the
kinematic factor for $++$ polarizations reads

{\footnotesize
\be
\begin{split}
\cK^{++} \sim &8\al^5 (s-2q\cdot q^*)(q\cdot p_4)^3(p_2\cdot\gz_3
p_3\cdot\gz_4 - p_3\cdot\gz_3 p_2\cdot\gz_4)z^3+\\
&4\al^5 (q\cdot
p_4)^2 \Bigg[ (s-2qq^*)(s+3t) (p_2\cdot\gz_3
p_3\cdot\gz_4 - p_2\cdot\gz_4 p_4\cdot\gz_3) \\
&-2sqq^*p_2\cdot\gz_3
p_3\cdot\gz_4 -s^2 p_2\cdot\gz_4 p_4\cdot\gz_3 + s u p_1\cdot\gz_3
p_2\cdot\gz_4 \\
&+ s t p_1\cdot\gz_4 p_2\cdot\gz_3 + s \gz_3\cdot\gz_4 (q\cdot p_4
q^*\cdot p_4 +\ov{4} tu ) \Bigg] z^2 +\cO(z)
\end{split}  \ .
\ee
}

\noindent
Letting $s-2q\cdot q^*=0$ the factor proportional to $z^3$ cancels and
the term in $z^2$ also gets simplified. The computation for the $--$
case is completely analogous.

{\footnotesize
\be
\begin{split}
\cK^{++} \sim &4\al^5 (q\cdot p_4)^2 \Bigg[ 4 p_3\cdot\gz_4
p_4\cdot\gz_3 + \gz_3 \gz_4 (2t+t^2-4q\cdot p_4 q^*\cdot p_4)
-4p_2\cdot\gz_3 p_3\cdot\gz_4\\
& +2(t+2)p_1\cdot\gz_3 p_2\cdot\gz_4
+ 2t p_1\cdot\gz_4 p_2\cdot\gz_3 +
4 p_1\cdot\gz_3 p_4\cdot\gz_3 \Bigg] z^2 + \cO(z)
\end{split}  \ ,
\ee

\be
\begin{split}
\cK^{--} \sim &4\al^4 (q\cdot p_4)^2 \Bigg[4 p_3\cdot\gz_4
p_4\cdot\gz_3 + \gz_3 \gz_4 (2t+t^2-4q\cdot p_4 q^*\cdot p_4)
-4p_2\cdot\gz_3 p_3\cdot\gz_4\\
& +2(t+2)p_1\cdot\gz_3 p_2\cdot\gz_4
+ 2t p_1\cdot\gz_4 p_2\cdot\gz_3 +
4 p_1\cdot\gz_3 p_4\cdot\gz_3 \Bigg] z^2 + \cO(z)
\end{split}  \ .
\ee
}

\noindent
Again, if we choose transverse polarization vectors for $\gz_1$ and
$\gz_2$ not to be the same, say along $T_1$ and $T_2$ ranging in the subset
of $d-4$ transverse directions, the kinematic factor is better behaved
by one power of $z$

{\footnotesize
\be
\cK^{TT} \sim -16 \al^4(\al s +1) z^4 (q\cdot p_4)^4
\gz_1^T\cdot\gz_2^T \gz_3\cdot\gz_4 + \cO(z^3)  \ ,
\ee

\be
\begin{split}
\cK^{T1T2} \sim &16 \al^4 (\al s +1) z^3 (q\cdot p_4)^3\Bigg(\gz_1^{T1}\cdot\gz_4
\gz_2^{T2}\cdot\gz_3+\gz_1^{T1}\cdot\gz_3 \gz_2^{T2}\cdot\gz_4\\
&+\al (p_4\cdot \gz_1^{T1} p_3\cdot \gz_1^{T2} +p_3\cdot
\gz_1^{T1} p_4\cdot \gz_1^{T2}) \gz_3\cdot\gz_4 \Bigg)
+\cO(z^2)
\end{split}  \ .
\ee
}

\section{Anti-symmetry of sub-leading terms in the superstring}
\label{app:antisymsuperstring}
In this appendix it will be shown that the tensor structure $B^2_{\mu\nu}$ in \eqref{eq:gluonssuperstringleadingz} is anti-symmetric in the Lorentz labels as advertised. This structure arises from the sub-leading contributions in $z$ in eq.~\eqref{eq:isolsusyamplead}. Although the leading terms are easy to isolate, the sub-leading ones require quite some more work. To start, note that any term proportional $\gz_1 \cdot \gz_n$ can be dropped at the outset as these can be absorbed immediately into an already known contribution. Both $\eta_n$ and $\theta_n$ need to be integrated over. Sub-leading ($\frac{1}{z}$ compared to leading) contributions in $\frac{1}{z}$ arise from terms with either $\theta_i$ and $\theta_j$ or $\eta_i$ and $\theta_j$, both for $i,j \neq n$. Terms with more $\theta's$ and $\eta's$ can also be dropped. To order the computation, note that four cases can be identified depending on whether $\eta_n$ and $\theta_n$ appear in either
\begin{multline}\label{eq:isolsysampspds}
I_1(\eta_n, \theta_n) \equiv \left[\sum_{j=2}^{n-1}\left( y_j (\gz_1 \cdot p_j) - \theta_j \eta_j (\gz_1 \cdot \gz_j)\right) \right. \\
\left. - \sum_{j,k=2}^n \left( \theta_j (\gz_1 \cdot p_j)  + \eta_j (\gz_1 \cdot \gz_j) \right) \left(  \eta_k (p_1\cdot\gz_k) +  (p_1 \cdot p_k) \gt_k \right)\right]  \ ,
\end{multline}
or
\begin{multline}\label{eq:isolsysampspds2}
I_2(\eta_n, \theta_n) \equiv \left[ \left( \prod_{1<j < n}\left(1- \frac{\gt_j \gt_n}{y_j}\right)^{p_j \cdot p_n} \right)   e^{\sum_{j=2}^{n-1} \frac{\eta_j(\gt_n-\gt_j) (p_n\cdot\gz_j) }{y_j}}  \right. \\
\left.\left( 1+ \sum_{i=2}^{n-1} \frac{\eta_n(\gt_i-\gt_n)(p_i\cdot\gz_n) - \eta_i \eta_n (\gz_n\cdot\gz_i)}{y_i - \gt_i \cdot \gt_n} \right)    \right]  \ .
\end{multline}
These contributions will be treated in turn.

\subsection*{Contribution of $ I_1\left(\eta_n, \theta_n\right)\, I_2\left(0,0\right)$}
The contribution of this part of the fermionic integration reads
\begin{equation}
\left[(\gz_1 \cdot p_n)(p_1\cdot\gz_n)  \right] \left[e^{- \sum_{j=2}^{n-1} \frac{\eta_j \gt_j (p_n\cdot\gz_j) }{y_j}} \right]  \ .
\end{equation}
Here all contributions proportional to the metric have been dropped as explained above. The order of the fermionic variables will be taken to be $\eta_n \theta_n$

\subsection*{Contribution of $ I_1\left(0,0\right)\, I_2\left(\eta_n, \theta_n\right)$}
The contribution of this part of the fermionic integration reads
\begin{multline}
\left[\sum_{j=2}^{n-1}\left( y_j (\gz_1 \cdot p_j) - \theta_j \eta_j (\gz_1 \cdot \gz_j)\right) \right. \\
\left. - \sum_{j,k=2}^{n-1} \left( \theta_j (\gz_1 \cdot p_j)  + \eta_j (\gz_1 \cdot \gz_j) \right) \left(  \eta_k (p_1\cdot\gz_k) +  (p_1 \cdot p_k) \gt_k \right)\right] \\
\left( e^{-\sum_{i=2}^{n-1} \frac{\eta_i\gt_i (p_n\cdot\gz_i) }{y_j}} \right)
 \left[\sum_{j=2}^{n-1} \left( \frac{  (p_j \cdot p_n) \gt_j +  \eta_j (p_n\cdot\gz_j)}{y_j}\right)\right.\\\left.
\left(\sum_{i=2}^{n-1} \frac{\gt_i(p_i\cdot\gz_n) +  \eta_i (\gz_n\cdot\gz_i)}{y_i} \right)
 -  \left(\sum_{i=2}^{n-1} \frac{(p_i\cdot\gz_n) + \eta_i \theta_i (\gz_n\cdot\gz_i)}{y_i} \right)    \right]  \ .
\end{multline}
The leading $z$ part of this contribution has no $\eta_i$ or $\theta_j$ for $i,j \neq n$.  The sub-leading terms of interest in this appendix have two of these. The exponential in both this and the previous contribution can be ignored, as the term proportional to $\sim \eta_i \theta_i$ will combine just as it did for the leading $z$ calculation. Furthermore, at leading order in $z$ $y_i = 1$, and this can be put in from this point on. This leaves
\begin{multline}\label{eq:tobecancellednonanti}
 \left(\sum_{j=2}^{n-1} (\gz_1 \cdot p_j) \right) \left(f(k_1, \gz_n) - \left[ \sum_{j=2}^{n-1} \theta_j \eta_j (\gz_n \cdot \gz_j)\right] \right) \\ - \left(\sum_{j=2}^{n-1} (p_j\cdot\gz_n)\right) \left(f(k_n, \gz_1)- \left[ \sum_{j=2}^{n-1} \theta_j \eta_j (\gz_1 \cdot \gz_j)\right] \right)  \ ,
\end{multline}
with
\begin{equation}\label{eq:thefunctionf}
f(k,\gz) =  - \sum_{j,k=2}^{n-1} \left( \theta_j (\gz \cdot p_j)  + \eta_j (\gz \cdot \gz_j) \right) \left(  \eta_k (p\cdot\gz_k) +  (p \cdot p_k) \gt_k \right)  \ .
\end{equation}
Hence it can be seen that at least part of this contribution is anti-symmetric in $\gz_1$ and $\gz_n$ at this order. The terms proportional to $f$ are not antisymmetric.

\subsection*{Contribution of $ I_1\left(0,\theta_n\right)\, I_2\left(\eta_n, 0\right)$}
The contribution of this part of the fermionic integration reads
\begin{multline}
\left[\left((\gz_1 \cdot p_n)\right) \sum_{k=2}^{n-1}  \left(\eta_k (p_1\cdot\gz_k) +  (p_1 \cdot p_k) \gt_k \right) \right. \\
\left.+   (p_1 \cdot p_n) \sum_{j=2}^{n-1} \left( \theta_j (\gz_1 \cdot p_j)  + \eta_j (\gz_1 \cdot \gz_j) \right)\right]\\
\left[- \sum_{i=2}^{n-1} \left(\theta_i (p_i\cdot\gz_n) + \eta_i (\gz_n\cdot\gz_i) \right) \right]  \ .
\end{multline}
The exponential can be ignored as any non-trivial contribution would involve at the least $4$ fermionic variables of particles $2$ to $n-1$, which would contribute at sub-sub-leading order in $z$. Likewise, all dependence on the variable $y$ can be dropped consistently at this order. Using momentum conservation and the function $f$ from eq.~\eqref{eq:thefunctionf} this can be written as
\begin{multline}
-\left[  (p_1 \cdot p_n) \sum_{j=2}^{n-1} \left( \theta_j (\gz_1 \cdot p_j)  + \eta_j (\gz_1 \cdot \gz_j) \right)\right]\left[ \sum_{i=2}^{n-1} \left(\theta_i (p_i\cdot\gz_n) + \eta_i (\gz_n\cdot\gz_i) \right) \right]\\
 -  \left(\sum_{j=2}^{n-1} (\gz_1 \cdot p_j) \right) f(k_1, \gz_n)  \ .
\end{multline}
The first term in this expression is antisymmetric in $\gz_1$ and $\gz_n$. The remaining term cancels part of the not anti-symmetric terms in the previous contribution in eq.~\eqref{eq:tobecancellednonanti}.

\subsection*{Contribution of $ I_1\left(\eta_n,0\right)\,I_2\left(0,\theta_n\right)$}
The contribution of this part of the fermionic integration reads
\begin{equation}
\left[\sum_{j=2}^n \left( \theta_j (\gz_1 \cdot p_j)  + \eta_j (\gz_1 \cdot \gz_j) \right) \left((p_1\cdot\gz_n) \right)\right]
\left[\sum_{k=2}^n \left(  \eta_k (p\cdot\gz_k) +  (p \cdot p_k) \gt_k \right) \right]  \ .
\end{equation}
A term proportional to the metric has been dropped here. The same remarks about exponential and $y_1$ apply as in the previous contribution. This term can be written as
\begin{equation}
-  \left(\sum_{j=2}^{n-1} (\gz_n \cdot p_j) \right) f(k_n, \gz_1)  \ ,
\end{equation}
which cancels the last not-anti-symmetric contribution in \eqref{eq:tobecancellednonanti}.

\subsection*{Conclusion}
Summing everything gives an anti-symmetric tensor contribution  at sub-leading order in large $z$. This is summarized by the anti-symmetric matrix $B_2$ in eq.~\eqref{eq:gluonssuperstringleadingz}.

\section{CSW rules for string amplitudes from a Risager shift?}
\label{app:CSWshifts} Since as shown in this article BCFW shifts
have an interpretation within string theory, a natural question is
if more general `shifts' of momenta yield similar results. In field
theory one example of this is known: the derivation of the CSW rules
for Yang-Mills theory \cite{Cachazo:2004kj} from a shift of $3$
particles \cite{Risager:2005vk}. In previous work
\cite{Boels:2008fc}, an analogue of these rules was derived for
Abelian string amplitudes from the DBI action in four dimensions. In
the Abelian case however, that result only resembles the CSW rules
superficially. In this appendix Risager's shift is applied directly
to the four-gluon superstring amplitude and it is shown that this at least
does not seem to yield a useful answer.

The color ordered four-point amplitude reads
\begin{equation}\label{eq:fourpoints}
A_4(1^-,2^-,3^+,4^+) = \frac{\langle 12 \rangle^4}{\langle 12
\rangle \langle 23 \rangle \langle 34 \rangle \langle 41 \rangle}
\frac{\Gamma(1 - \al s) \Gamma(1 - \al t)}{\Gamma(1 + \al u)} \ .
\end{equation}
Risager's shift was designed to leave the MHV amplitude invariant.
Since this amplitude contains holomorphic spinor products only, only
anti-holomorphic on-shell spinors should be shifted such that
momentum conservation is conserved. This is possible for a three
particle shift,
\begin{equation}
\left\{ \begin{array}{ccc}
1_{\alpha} &\rightarrow & 1_{\alpha} + z \eta_\alpha \braket{2 3} \\
2_{\alpha} &\rightarrow & 2_{\alpha} - z \eta_\alpha \braket{1 3} \\
3_{\alpha} &\rightarrow & 3_{\alpha} + z \eta_\alpha \braket{1 2}
\end{array} \right.  \ ,
\end{equation}
with an arbitrary $\eta$. This shifts the Mandelstam variables as
\begin{equation}
\left\{ \begin{array}{ccc}
s &\rightarrow & s + z \sbraket{\eta 4} \braket{3 4}\braket{1 2} \\
t &\rightarrow & t + z \sbraket{\eta 4} \braket{2 3}\braket{1 4}  \\
u &\rightarrow & u - z \sbraket{\eta 4} \braket{1 3}\braket{2 4}
\end{array} \right.  \ .
\end{equation}
Note that the choice $\eta = 4_{\alpha}$ is a special case: for this
value of $\eta$ the shift leaves the full amplitude manifestly
invariant. In all other cases, the amplitude can be evaluated as a
contour integral
\begin{equation}
A(0) = \int_{z=0} \frac{A(z)}{z} \ .
\end{equation}
The right hand side of this equation does not depend on $\eta$ after
a change of variables $z' = \sbraket{\eta 4} z$,
\begin{equation}
A(0)=  \frac{\langle 12 \rangle^4}{\langle 12 \rangle \langle 23
\rangle \langle 34 \rangle \langle 41 \rangle} \int_{z=0}
\frac{1}{z} \left(\frac{\Gamma(1 - \al s - z' \al \braket{3
4}\braket{1 2}) \Gamma(1 - \al t - z' \al \braket{2 3}\braket{1 4}
)}{\Gamma(1 + \al u - z' \al \braket{1 3}\braket{2 4}  )} \right) \
.
\end{equation}
Pulling the contour to infinity yields a double infinite sum over
the residues and a possible contribution from the contour at
infinity,
\begin{multline}
A(0) = A(\infty) + A_{\textrm{tree}} \sum_{i=0}^{\infty}
\frac{(-1)^i}{\Gamma(i+1)} \left(\frac{1}{1 - \al s +i}
\frac{\Gamma\left(1-\al t - (i + 1 - \al s)\frac{\braket{2
3}\braket{1 4}}{\braket{3 4}\braket{1 2}} \right)}{\Gamma\left(-i -
\al t - (i + 1 - \al s)\frac{\braket{2 3}\braket{1 4}}{\braket{3
4}\braket{1 2}} \right) } + \right. \\ \left.  \frac{1}{1 - \al t
+i}  \frac{\Gamma\left(1-\al s - (i + 1 - \al t)\frac{\braket{3
4}\braket{1 2}}{\braket{2 3}\braket{1 4}} \right)}{\Gamma\left(-i -
\al s - (i + 1 - \al t)\frac{\braket{3 4}\braket{1 2}}{\braket{2
3}\braket{1 4}} \right) } \right) \ .
\end{multline}
The resulting sum is reminiscent of a non-adjacent BCFW shift up to
the phase factors which permeate the expression. Indeed, if this
factor is equal to the imaginary number
\begin{equation}
\frac{\braket{3 4}\braket{1 2}}{\braket{2 3}\braket{1 4}} = \ii \ ,
\end{equation}
the infinite sum equals the original amplitude. In all other cases
we were unable to interpret or sum the infinite sum. An $\al$
expansion of the result also does not seem to converge. If it is true that the
infinite sum equals the Beta function inside \eqref{eq:fourpoints},
this constitutes a functional identity which, as far as we are
aware, is unknown and we suspect is wrong.

\addcontentsline{toc}{section}{References}

\bibliographystyle{JHEP}
\bibliography{recurbiblio}

\end{document}